\documentclass[12pt]{article}
\usepackage{amsmath,amsthm,amssymb,euscript,epsfig,youngtab}
\usepackage{array}
\usepackage{fancybox}
\usepackage[usenames]{color}
\usepackage{amsfonts,bm}
\usepackage{graphicx}

\newcommand{\be}{\begin{equation}}
\newcommand{\ee}{\end{equation}}
\newcommand{\bea}{\begin{eqnarray}\displaystyle}
\newcommand{\eea}{\end{eqnarray}}

\makeatletter
\@addtoreset{equation}{section}
\makeatother
\renewcommand{\theequation}{\thesection.\arabic{equation}}
\def\one{{\hbox{ 1\kern-.8mm l}}}
\def\zero{{\hbox{ 0\kern-1.5mm 0}}}

\def\cM{{\cal M}} \def\cN{{\cal N}} \def\cO{{\cal O}}

 \def\cZ{{\cal Z}}

\def\s{ \sigma }

\def\mZ{ \mathbb{Z} }

\def\mR{ \mathbb{R}}

\begin{document}

\makeatletter
\@addtoreset{equation}{section}
\makeatother
\renewcommand{\theequation}{\thesection.\arabic{equation}}

\rightline{QMUL-PH-11-16}
\rightline{WITS-CTP-080}
\vspace{1.8truecm}

\vspace{15pt}


{\LARGE{  
\centerline{   \bf Strings from Feynman Graph counting : } 
\centerline{ \bf  without large N }
}}  

\vskip.5cm 

\thispagestyle{empty} \centerline{
    {\large \bf Robert de Mello Koch
${}^{a,} $\footnote{ {\tt robert@neo.phys.wits.ac.za}}}
   {\large \bf and Sanjaye Ramgoolam
               ${}^{b,}$\footnote{ {\tt s.ramgoolam@qmul.ac.uk}} }
                                                       }

\vspace{.4cm}
\centerline{{\it ${}^a$ National Institute for Theoretical Physics ,}}
\centerline{{\it Department of Physics and Centre for Theoretical Physics }}
\centerline{{\it University of Witwatersrand, Wits, 2050, } }
\centerline{{\it South Africa } }

\vspace{.4cm}
\centerline{{\it ${}^b$ Centre for Research in String Theory, Department of Physics},}
\centerline{{ \it Queen Mary University of London},} \centerline{{\it
    Mile End Road, London E1 4NS, UK}}

\vspace{1.4truecm}

\thispagestyle{empty}

\centerline{\bf ABSTRACT}

\vskip.4cm 

A well-known connection between $n$ strings winding around a circle 
and permutations of $n$ objects plays a fundamental role in the string theory 
of large $N$ two dimensional Yang Mills theory and elsewhere in 
topological and physical string theories.  Basic questions 
in the enumeration of Feynman graphs can be expressed 
elegantly in terms of permutation groups. 
We show that these permutation techniques for Feynman 
graph enumeration, along with the Burnside counting lemma, 
lead to  equalities between counting problems of Feynman graphs 
in scalar field theories  and Quantum Electrodynamics
with  the counting of amplitudes in a string theory with torus or cylinder 
target space. This string theory arises in 
the large N expansion of two dimensional Yang Mills and is closely related 
to lattice gauge theory with $S_n$ gauge group.
We collect and extend results on generating functions for Feynman graph 
counting, which connect directly with the string picture. 
We propose that  the connection between string combinatorics 
and permutations has implications for QFT-string dualities,  beyond 
the framework of large $N$ gauge theory.

\setcounter{page}{0}
\setcounter{tocdepth}{2}

\newpage

\tableofcontents

\setcounter{footnote}{0}

\linespread{1.1}
\parskip 4pt

{}~
{}~

\newpage

\section{ Introduction }

There is a basic connection between winding states of strings
and partitions of numbers. If we have a string wound around a circle, it 
can have a winding number which is a positive integer. For multiple 
strings, the winding numbers can be added to define a total winding number.  
In the sector of the multi-string Hilbert space with total winding 
number $n$, the partitions of $n$, denoted $p(n) $, correspond to 
the different types of string states. The number 
$p(n)$ also plays an important role in connection with the 
symmetric group $S_n$, of order $n!$,  consisting of the re-arrangements of 
the integers $\{ 1 , 2 \cdots , n \} $. It is the number of 
conjugacy classes in $S_n$.

In generic string theories, a string state is characterized by
a winding number along with additional vibrational and momentum 
quantum numbers. In simple cases, the winding numbers completely characterize 
the string state. A particularly striking example of this simplification 
occurs in  the string dual of 
two dimensional Yang Mills theory (2dYM) on a Riemann surface $ \Sigma_{ G  } $ 
with  genus $G$.  This theory, for 
$U(N)$ gauge group, can be solved exactly \cite{migdal}.  
The coefficients in the large $N$ expansion of amplitudes 
in 2dYM
defined on Riemann surfaces can be expressed in terms of 
sums over symmetric group elements. This combinatoric data 
has an interpretation in terms of branched covers of $ \Sigma_{ G } $
by a string worldsheets $\Sigma_h$,  Riemann surfaces
of genus $h$ related to the 
order in the $1/N$ expansion \cite{gross1992,gt1,gt2}.
This provides a beautiful realization of the ideas of \cite{tHooftPlanar}.  
These spaces of covers are also called  Hurwitz spaces and
they are spaces of holomorphic maps from $\Sigma_h $ to $\Sigma_G$.  
These spaces are reviewed from the point of view of 2dYM in \cite{CMR}.

Of particular interest 
are manifolds $ \Sigma_{ G , B  } $, with genus $G$ and $B$ boundaries.  
The observables of 2dYM are defined as functions of 
the boundary holonomies taking values in the gauge group $U(N)$. 
The gauge-invariant functions of these group elements are 
multi-traces, which are also classified 
for fixed degree $n$, by partitions. This leads to an interpretation of 
the boundary partition functions in terms of covering spaces of 
the manifold with boundary.  Generically, the partition function sums over  
branched covers. The derivative of the map at some points 
on the worldsheet can vanish, and the images of such points 
are called {\it branch points}. 
 A particularly simple situation occurs when the 
Yang Mills theory is defined on a cylinder, and the area is taken to 
zero. Then the partition function sums only over {\it unbranched covers}.

The large $N$ expansion of 2dYM can also be expressed 
in terms of lattice gauge theory with $S_n$ gauge group
which can be formulated following Wilson and letting edge holonomies 
take values in $S_n$, with an appropriate plaquette action. 
This theory is topological and will be called $S_n$ TFT (topological field theory). So along with the original $U(N)$ description, there is 
the picture of $S_n$ TFT, as well as the Hurwitz space description.
 The Hurwitz space 
connects most directly to the equations formulated on the string worldsheet, 
since the holomorphy condition is the localization locus 
for the string path integral. 
Explicit worldshseet actions have  been proposed \cite{CMR,horava,vafa,AOSV,szabo2dYM}. The 2d $S_n$ TFT  can be viewed
as the string field theory of this Hurwitz space string theory. 
 In this paper we will find the $2d$ $S_n$ TFT particularly useful.

Other examples where the string-permutation connection is important is
Matrix string theory\cite{dvv}. An  orbifold superconformal field theory
with  target space $S^N ( \mR^8 )$ arises as an IR limit of 
$N=8$ SYM in two dimensions. Conjugacy classes  in $S_N $ 
correspond to string winding numbers which are dual to momenta 
of gravitons in the Matrix Theory interpretation.  
Yet another example is the connection between Belyi maps 
(a special class of branched covers with sphere target space)
and correlators of the Gaussian Hermitian matrix model \cite{BZ,dMRam}.

From the above we take the lesson that the connection 
between permutations and strings is rather generic. 
In many cases, it is a  tool to implement 
the ideas of strings emerging from the  large $N$ expansion
in quantum field theories. Now a large number of counting 
problems can be formulated in terms of permutations acting 
on various types of sets. Some of these counting 
problems, such as the ones relevant to 2d Yang Mills 
and to Belyi maps, involve permutations acting on each other, 
for example by group multiplication. It will not surprise 
anybody that the question of counting Feynman graphs in quantum field
theories, {\it without large $N$}, can be formulated in terms of
symmetric groups acting on some sets having to do with vertices and edges.

Investigating the counting of Feynman graphs more carefully reveals
that, in fact, these counting problems can be formulated 
in terms of rather intrinsic 
properties of the symmetric groups themselves. The nature of the 
specific interactions determine the form of the vertices, and in turn 
certain {\it subgroups}  of symmetric groups, which can be viewed  as 
the symmetry of all the vertices. 
Then  we draw in an idea from the study of Belyi maps and 
their associated ribbon graphs, which is called {\it cleaning}. 
This is the construction of associating to a given graph, a new graph 
obtained by introducing a new type of vertex in the middle 
of the existing edges, dividing them into half-edges.
 This allows the description of Wick contractions 
in terms of permutations which are pairings. This was exploited 
recently in a physics setting in \cite{dMRam}.

This line of thinking culminates in an elegant way 
to compute numbers of Feynman graphs of real scalar fields,  
which coincides with the results in a classic in the mathematical 
literature on  graph counting  \cite{Read}. Our derivation does not rely on 
the notion of graph superposition used in  \cite{Read} 
and appears more direct, at least when approaching these 
graphs from the perspective of Wick contractions in perturbative 
quantum field theory. We also show that the formulae counting 
Feynman graphs, as well as those describing symmetry factors of individual 
graphs, can be interpreted as observables in the 2d  $S_n$  TFT. 
 The number $n$ will depend on the number of vertices 
and edges in the Feynman graph calculation. 
We extend our considerations to QED, deriving new counting results
on Feynman graphs. 
Again we obtain an expression of Feynman graph counting problems in terms of 
observables in the 2d $S_n$ TFT.  An interesting aside 
is a surprising  connection between the counting of QED Feynman graphs 
and the counting of ribbon graphs. 

While the string-permutation connections 
underlie the emergence of strings from large $N$ 
2d YM, our results suggest that these same connections 
could also lead to the emergence of strings from 
quantum field theories such as that of a scalar field or QED.

We now give an outline of the paper. 
In section 2, we review the connection between strings and permutations, 
which plays a prominent role in the string theory of large $N$ 
2d Yang Mills theory with $U(N)$ gauge group. 
We describe the perspective of lattice gauge theory with $S_n$ 
gauge group and explain its topological nature by drawing on 
existing 2dYM literature.  We also review the Hermitian matrix model and its connection to Belyi maps, as another realization of the string-permutations
connection.  Finally, we mention the Burnside Lemma from combinatorics, 
which is used very generally for counting orbits of group actions.  
In section 3, we review Feynman rules for $\phi^4$ theory. 
We recall how the symmetry factors get multiplied with additional group 
theory factors associated with global or gauge symmetry groups, 
and space-time integrals. Our main  focus will be the combinatorics of 
the graphs and their symmetry factors.

In section \ref{FeynPair},
 we describe a method of enumerating Feynman graphs and
calculating their symmetry factors, which relies on a pair of 
combinatoric objects. The first of the pair, which we call $\Sigma_0$, 
captures the form of the vertices. The second piece of data, which 
we call $ \Sigma_1$, is associated with the Wick contractions. 
Given a graph, this data can be constructed by putting a new vertex
in the middle of each edge. We can imagine these vertices to be coloured 
differently from the ones already present in the graph. This splits each 
existing edge into a pair of half-edges. We associate labels $\{ 1 , 2, \cdots \}$ 
with each of these half-edges. This is explained in the context of real 
scalar field theory, in the first instance, with vacuum graphs in
 $ \phi^4$ theory.  This formulation leads to simple
 one-liners  in GAP (mathematical 
software of choice for group theory) \cite{GAP} for calculations 
of symmetry factors and enumeration. This is explained with examples 
in Appendix 
\ref{FeynGAP} 

In section \ref{sec:string}, we  use the Burnside Lemma
to count Feynman graphs, and obtain the first hints of stringy 
geometry of maps to a torus. The picture of maps to a torus 
has some intricacies, and it turns out that a deeper 
understanding of the combinatorics leads to a simpler picture
in terms of strings covering a cylinder. This is is achieved 
by first recalling from the maths literature that graph counting 
formulae are expressed in terms of  a certain double coset 
\cite{Read}. 
We find that the formulation of graph counting in terms of 
the pair   $ ( \Sigma_0 , \Sigma_1 ) $  finds a natural meaning 
in terms of the double coset.  This allows us to  exhibit the counting 
of Feynman graphs in scalar field theory 
as an observable in 2d $S_n$ TFT on a cylinder. Following standard 
constructions in covering space theory, the $S_n$ data  is used 
to  construct unbranched covers of the cylinder. Some of the counting 
formulae that follow from the double coset picture can be understood 
using the idea of Fourier transformation on symmetric groups. Calculations
explaining this are in Appendix  \ref{functions on double coset}.   
Interestingly very similar calculations are relevant to recent 
results in correlators of BPS operators in $N=4$ super-Yang-Mills 
\cite{BHR}.

We  pursue, following \cite{Read},  the application of the double coset 
picture to derive generating functions for graph counting in section
\ref{sec:numfeyn}. We also explain here how the formulae change 
when we generalize our considerations from $ \phi^4$ to 
$\phi^3$ and other interactions. Generalizations away from 
vacuum Feynman graphs to the case with external edges is also 
explained. The double coset formulae lead most directly to 
the set of all Feynman graphs, included disconnected ones. 
Counting formulae for the connected ones are obtained here. 
Appendix \ref{data} contains explicit lists for Feynman graph counting. 
The first few terms agree with existing physics literature. 
A few of the series we consider are listed in the Online 
Encyclopaedia of Integer sequences \cite{OEIS}, but the majority are not 
listed there.

In section \ref{sec:ribbon}, we show that these approaches work in a simple way 
for the ribbon graphs which arise in 
doing the $1/N$ expansion. While the ribbon graphs for large $N$ 
are traditionally drawn using double lines, they can also be defined 
in a way closer to ordinary graphs, with single lines but with 
the difference that the vertices have a cylic order (see e.g. 
\cite{Looijenga} \cite{lanzvon}).  
This allows cyclic symmetry of the half-edges at 
the vertices in testing equivalence of differently labelled
ribbon graphs, but not arbitrary permutations. 
Taking this into account, we express the question of 
how many Feynman diagrams correspond to the same ribbon graph 
(embedded graph) in group theoretic terms, associated with the 
symmetry breaking from permutation group to cyclic group.

In section \ref{sec:QEDYUK}, we apply  these ideas to QED or 
Yukawa theory, giving the connection to 2d $S_n$ TFT 
and deriving generating functions. 
In section \ref{sec:QEDfurry}, we adapt the counting to QED, with 
the vanishings due to Furry's theorem taken into account.  
Again we obtain the 2d  $S_n$ TFT  connection and the generating functions. 
Manipulations of the double coset description for QED 
leads to a somewhat surprising connection between QED graph
counting and ribbon graph counting. We explain this 
by describing a bijection between these graphs. The arrows 
on electron loops provide the cyclic symmetry which is 
key to ribbon graphs. A consequence is that 
the number of QED/Yukawa vacuum 
graphs with $2v$ vertices is equal to the number of ribbon graphs with $v$
edges.

In section \ref{sec:discussion}, 
we discuss our results with a view to extracting some  
lessons for gauge string 
duality. Ribbon graphs give a vivid physical picture of how strings arise
from quantum field theory, but an equally compelling and arguably simpler 
physical picture is that of strings winding on a circle being 
described by permutations.  
This latter picture has been exploited here to give new connections 
between QFT Feynman diagrams and string counting. The natural question
is whether this connection can be extended to extract from 
Feynman graphs of QFT, without large $N$, 
something more than stringy combinatorics to include space-time 
dependences of correlators and S-matrices.

In section \ref{sec:summout}, 
we summarize the paper and discuss some concrete avenues for  future work. 
Appendix \ref{nutshell} describes a semi-direct product 
structure of the Automorphism groups, whose orders give symmetry factors 
of Feynman graphs, and points out a difference between the 
notion of Automorphism most commonly used in the mathematics literature 
on graphs and the one relevant to symnmetry factors of Feynman graphs.

\section{ Review of strings and permutations }\label{revstringsperms}

Consider a string wrapping a circle. The winding number is a topological 
characteristic of the map. In string theory, the string is part of the worldsheet, 
the circle is viewed as part of target space. Let the target circle be 
parametrized by $X$ with $ X \sim X + 2 \pi $. Let the string be parametrized 
by $ \sigma $ with $ \sigma \sim \sigma + 2 \pi $. A string with winding number $k$ 
is described by 
\bea 
X = k \sigma 
\eea
For multiple strings, we may have distinct winding numbers. Adding the winding numbers
gives the degree of the map from the strings. 

\begin{figure}[ht]
\begin{center}
\resizebox{!}{3.5cm}{\includegraphics{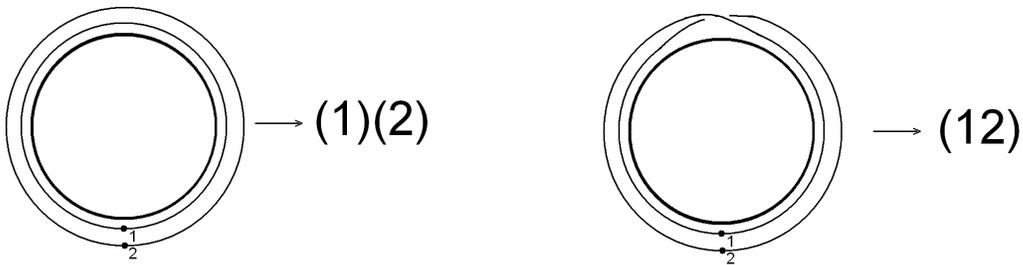}}
\caption{ Winding strings and permutations  }
\label{fig:wrap}
\end{center}
\end{figure}

Given such a configuration of winding strings, 
we may label the inverse images of a fixed 
point on the circle from $1$ to $n$. 
Following the inverse images as we move round  the target circle
yields a permutation. 
The two possible winding configurations at $n=2$ are shown in 
Figure \ref{fig:wrap}.

In ordinary string theory, the winding number is 
one of the quantum numbers specifying 
a string state. In some topological settings, the string winding number is the
 only relevant quantum number.

\subsection{ Review of results from 2d Yang Mills }\label{2dYM}

\begin{figure}[ht]
\begin{center}
 \resizebox{!}{4cm}{\includegraphics{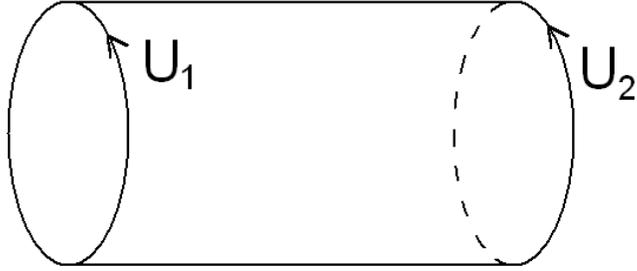}}
\caption{ The 2dYM partition function with boundary holonomies is known exactly }
 \label{fig:cylinder}
\end{center}
\end{figure}

Recall that 2d Yang Mills on a target space with a boundary has the partition function 
\bea 
Z = \sum_R ( {\rm Dim} R )^{ 2 - 2G - B }  e^{ - A C_2 ( R ) }  \prod_i \chi_R ( U_i)  
\eea
$G$ is the genus of the surface, $B$ is the number of boundaries, 
$U_i$ are the holonomies 
at the boundaries and $A$ is the area of the target and $C_2(R)$ 
is the quadratic Casimir \cite{migdal}.
The choice of fixed boundary holonomies is illustrated in Figure \ref{fig:cylinder}.

For a target cylinder, we have 
\bea 
Z ( U_1 , U_2 ) = \sum_R \chi_R ( U_1 ) \chi_R  ( U_2) e^{ - A C_2( R ) }  
\eea
in terms of characters of $U(N)$ elements. 
The power of the dimension is zero because $G = 0 , B=2 $. 
The characters have an expansion in multi-traces. These traces can be constructed 
from permutations as follows 
\bea 
tr ( \sigma U ) \equiv  U^{i_1}_{i_{\sigma (1)}  } U^{i_1}_{i_{ \sigma (2 )} }
 \cdots U^{i_1}_{i_{ \sigma (n ) } } 
\eea
This is a very useful formula when developing a string interpretation for Wilson loops in 2dYM \cite{ramwil}. 

Using these observables for the cylinder, we define 
\bea 
Z ( \sigma_1 , \sigma_2 ) = \int dU_1 dU_2 tr ( \sigma_1 U_1^{\dagger}  ) tr ( \sigma_2 U_2^{\dagger} ) 
  Z ( U_1 , U_2) 
\eea
This integral can be done, and in the zero area limit $A=0$ we find 
\bea 
Z ( \sigma_1 , \sigma_2 ) = \sum_{ R \vdash n } \chi_R ( \sigma_1 ) \chi_R  ( \sigma_2 ) 
\eea
The notation $R  \vdash n $ means that $R$ is running over partitions
of $n$ (row lengths of Young diagrams) which parametrize irreducible 
representations of $S_n$, and the $\chi_R(\sigma)$ are characters 
of $S_n$ elements. 
Using the invariance of the character under conjugation, we can replace 
\bea  
\chi_R ( \sigma_2 ) =  
{ 1 \over n ! } \sum_{ \gamma \in S_n } \chi_R ( \gamma \sigma_2 \gamma^{-1})
\eea
This gives 
\bea\label{Zsig1sig2} 
Z ( \sigma_1 , \sigma_2 ) = { 1 \over n! } \sum_{ \gamma \in S_n } 
\sum_{ R \vdash n } \chi_R ( \sigma_1 ) \chi_R  ( \gamma \sigma_2 
 \gamma^{-1} ) 
\eea
Since $\sum_{ \gamma }  \gamma \sigma_2 \gamma^{-1} $ 
is a central element in the group algebra of $S_n$, 
we can use Schur's Lemma to combine the product of
characters into a single character 
\bea 
Z ( \sigma_1 , \sigma_2 ) = { 1 \over n! } \sum_{ \gamma \in S_n } 
\sum_{ R \vdash n } d_R \chi_R ( \sigma_1  \gamma \sigma_2 
 \gamma^{-1} ) 
\eea
Now use the Fourier expansion on the symmetric group, which 
allows the delta function to be written in terms of characters, to 
obtain 
\bea 
Z ( \sigma_1 , \sigma_2 ) = \sum_{ \gamma } \delta ( \sigma_1 \gamma \sigma_2 \gamma^{-1} ) 
\label{deltasum}
\eea

When the gauge group is a product $U(N) \times U(N) \times U(N)$, we have to specify a 
holonomy for each gauge group in the product. Denote them by $U,V,W$. Recalling that the
character of a direct product is the product of the characters we have 
the zero area partition function for a cylinder 
\bea 
Z ( U_1, V_1 , W_1 ; U_2  , V_2 , W_2 ) = 
\sum_{ R, S , T } \chi_{R} (U_1)  \chi_{S } (V_1)  \chi_{ T } (W_1)  \chi_{R} (U_2)  \chi_{S } (V_2)  \chi_{ T } ( W_2 ) ) 
\eea
The boundary observables in this case, in a basis appropriate 
for a string interpretation, 
 are $ tr ( \sigma U )  tr (  \rho V ) tr ( \tau W )  $.  
In terms of permutations 
\bea 
Z ( \sigma_1, \rho_1 , \tau_1 ; \sigma_2, \rho_2 , \tau_2) 
= \sum_{ \gamma \in S_{n } \times S_n \times S_n } 
  \delta (   \sigma_1 \circ  \rho_1 \circ  \tau_1 ~ \gamma ~ 
 \sigma_2 \circ  \rho_2 \circ  \tau_2 ~ \gamma^{-1} ) 
 \eea

We have  discussed above the classification  of string maps to 
a circle target space, in motivating the choice of boundary observables 
in large $N$ 2dYM. The  full interpretation of this result requires
an extension of our considerations to a   Riemann surface target. 
In this case, 
the strings-permutations connection 
has deeper implications captured in the  { \it Riemann existence theorem}.  
Consider the equivalence classes of holomorphic maps from the worldsheet ($\Sigma_h$) to the target ($\Sigma_G$)
with two maps $f_1$ and $f_2$ defined to be equivalent if these exists a 
holomorphic invertible map $\phi:\Sigma_h\to\Sigma_G$
such that $f_1=f_2\circ \phi$. Given a holomorphic map (branched cover) with $L$ branch points and of degree $n$, 
we can  obtain a combinatoric description 
by picking a generic base point and labeling the inverse images as  
$ \{ 1, 2, \cdots , n \} $.  By following the inverse images of a 
closed path starting at the base point and encircling 
each branch point, we can get a sequence $\sigma_1,\sigma_2,...,\sigma_L$ 
of permutations. For $G >0$, there are also permutations for the 
$a,b$ cycles of $\Sigma_G$, denoted $s_i,t_i$ for $i=1 \cdots G$.
 Two equivalent holomorphic maps are described by permutations $\sigma_1,\sigma_2,...,\sigma_L, s_1,t_1 , \cdots , s_G,t_G $ 
and $\sigma'_1,\sigma'_2,...,\sigma'_L, s_1',t_1' , \cdots , s_G',t_G'$ 
which are related by conjugation $\sigma_i=\alpha\sigma'_i \alpha^{-1} ,
s_i= \alpha s'_i\alpha^{-1} , t_i=\alpha t'_i\alpha^{-1} $. 
This correspondence between sequences of permutations and holomorphic
maps is captured in the Riemann existence theorem (see 
for example \cite{lanzvon}).  
Relations in the fundamental group of the Riemann surface 
punctured at the branch points are reflected in the 
sequence of permutations. 

The delta function in  (\ref{deltasum}) enforces a
relation in the fundamental group of the cylinder.
When this partition function is generalized beyond zero area, 
the sum (\ref{deltasum}) is modified to include additional
permutations which can be interpreted as a counting of branched covers.
Even at zero area, but for other Riemann surfaces, there are 
additional permutations associated with branch points. 
We will come back to this 
point in the discussion of (\ref{3bdpart}).

\subsection{ Lattice topological field  theory for $S_n$ } 

\begin{figure}[ht]
\begin{center}
 \resizebox{!}{4.5cm}{\includegraphics{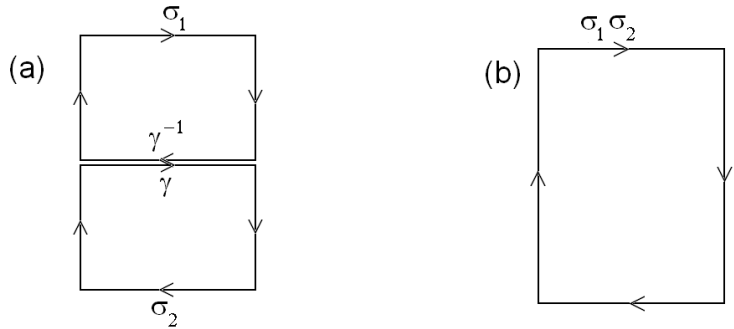}}
\caption{Two discretizations of the disc with the same boundary condition.}
 \label{fig:latticefieldtheory}
\end{center}
\end{figure}

In lattice gauge theory\cite{Wilson} for two dimensions, 
we discretize (e.g triangulate or give a more general cell decomposition for)
the Riemann surface. To each edge we associate a 
group element. To each plaquette, we associate a 
weight which depends on the product of group elements
along the boundary of the plaquette. Let us choose the plaquette weight to be
\bea 
Z_{P} ( \sigma ) = \delta ( \sigma ) = \sum_{ R \vdash n } { d_R\over n! } 
 \chi_R ( \sigma ) 
\eea
where $ \sigma $ is the product of group elements around the boundary of 
the plaquette. 
The partition function for the discretized manifold 
is the product of weights for all the plaquettes. 
The partition function can be shown to be invariant 
under refinement of the lattice so that the result depends only on the 
topology of the Riemann surface.  This 
is proved in \cite{wittenQuantum2d}. The language 
used there is that of continuous groups, but by replacing 
integrals with sums, it applies equally well to finite groups.
In \cite{wittenQuantum2d} surfaces with a finite area $A$ are considered.
For the topological aspects of the theory that interest us, it is
enough to consider $A=0$.
For manifolds with boundary one fixes the group elements 
at the boundaries. As a simple example, consider a lattice theory defined 
on a disc. The target might be discretized with either one plaquette or 
two plaquettes as shown in Figure \ref{fig:latticefieldtheory}.
Starting from the discretization given in (a) we find
\bea
Z (\sigma_1 , \sigma_2 ) &=&   \sum_{ \gamma }Z_{P} ( \sigma_1\gamma^{-1} )Z_{P} ( \sigma\gamma )\nonumber\\
&=& \sum_{ \gamma } \delta ( \sigma_1 \gamma^{-1})\delta(\sigma_2 \gamma)\nonumber\\
&=&\delta (\sigma_1\sigma_2)
\eea
which is precisely the partition function obtained using the
discretization shown in (b). The generalization to finer discretization
works in exactly the same way.

For the cylinder  
\bea 
Z (\sigma_1 , \sigma_2 ) =   \sum_{ \gamma } \delta ( \sigma_1 \gamma \sigma_2 \gamma^{-1} ) 
\eea
which is of course (\ref{deltasum}). 
For a 3-holed sphere, 
\bea\label{3bdpart}  
Z (\sigma_1 , \sigma_2 , \sigma_3 )
&& ={ 1 \over (n!)^2  }  
\sum_{ \gamma_1 , \gamma_2 } \delta ( \sigma_1 \gamma_2 \sigma_2 \gamma_1^{-1} 
\gamma_2 \sigma_3 \gamma_2^{-1}  ) \cr 
&& = {1 \over n! } \sum_{ R }
   { \chi_R ( \s_1 )  \chi_R ( \s_2 )  \chi_R ( \s_3 )\over d_R } 
\eea
This is also derived in \cite{FHK} from a more axiomatic approach. 

These formulae from lattice gauge theory of $S_n$ arise 
from the leading the large $N$ limit of the 
$U(N) $ gauge theory at zero area. The cylinder case is special in that 
this is the full answer to all orders in the $1/N$ expansion. The answer 
in (\ref{3bdpart})  is obtained after using the leading large 
approximation of $\Omega $, which is a sum over 
elements in $S_n$ weighted by powers of $N$.  
See for example 6.1.2 of \cite{CMR} for a discussion of $\Omega$.
At subleading orders, 
the $\Omega $ factor contains a sum over permutations, which can be interpreted
in terms of  sources of curvature in the $S_n$ bundle.
From the Hurwitz space string interpretation of 2dYM, there are  branch points 
in the interior. 

The $S_n$ lattice gauge theory perspective for $U(N)$ gauge theory 
at large $N$ is emphasized in \cite{dada}. Computations of 2dYM 
partition functions  for general Riemann surfaces with boundaries 
expressed in the symmetric group basis, and the connection 
with Hurwitz space counting is given in \cite{gt2,CMR,ramwil}.
The connection to topological field theory with $S_n$ gauge group 
was also observed in \cite{dijkME}.

To summarize, the large $N$ expansion of 2dYM with $U(N)$ gauge group 
on a Riemann surface $ \Sigma_{ G , B } $ (genus $G$, with $B$ boundaries)
can be expressed in terms of symmetric groups. This combinatoric 
data arising has an interpretation as 2d gauge theory with $S_n$ gauge group.
This connects directly with the string theory interpretation 
in terms of maps from a string worldsheet $ \Sigma_h $ 
(genus $h$ related to the order in the $1/N$ expansion) using 
classic results relating the space of branched covers (Hurwitz space) 
to symmetric groups. 

We will find in this paper that counting problems 
of Feynman graphs can be expressed in terms of 
 certain generalizations of the 2dYM results, which 
can be expressed elegantly in terms of the 2d 
 $S_n$ gauge theory perspective, and also admit a connection 
to branched covers.

\subsection{ Permutations and Strings beyond 2dYM }

The connection between strings and permutations also has
 interesting implications
for the correlators of the Gaussian hermitian one-Matrix model \cite{dMRam}.  
The computation of correlators can be mapped to
 the counting of certain triples of 
permutations which multiply to the identity. These count
 holomorphic maps from
 world-sheet to sphere target with three branch points on the
target. Holomorphic maps with three branch points are related, by
Belyi's theorem, to curves and maps defined over algebraic numbers. 
Thus, the string theory dual of the one-matrix model at generic couplings has worldsheets defined
over the algebraic numbers and a sphere target. For related ideas see \cite{newtom,rajesh}.

Finally, yet another connection between strings and permutations that is relevant to our
present discussion is provided by Matrix String Theory, defined by the
IR limit of two-dimensional $N=8$ SYM. This limit is strongly coupled and
a nontrivial conformal field theory describes the IR fixed point.
The conformal field theory is the $N = 8$ supersymmetric sigma model on the orbifold target
space $(\mR^8)^N/S_N$, formed from the eigenvalues of the Higgs fields $X_I$ of the theory \cite{dvv}. 
If we go around the space-like $S^1$ of the world-sheet, the eigenvalues can be interchanged.
Concretely
$$
    X_I(\sigma + 2\pi) = X_{g(I)}(\sigma)
$$
where $g$ takes value in the Weyl group of $U(N)$ which is the symmetric group
$S_N$. These twisted sectors correspond to configurations of strings with
various lengths. Consequently, twisted sectors with given winding number are
labeled by the conjugacy classes of the orbifold group $S_N$.

\subsection{ A useful theorem in combinatorics} 

Let $G$  be a finite group that acts on a set $X$. For each $g$  in $G$  let $X^g$ denote the set 
of elements in $X$  that are fixed by $g$. Burnside's lemma asserts the following formula for the 
number of orbits 
\bea
   \hbox{ Number of orbits of the $G$-action on $X$}  = \frac{1}{|G|}\sum_{g \in G}|X^g|.
\eea
Thus the number of orbits  is equal to the average number of points fixed by an element of $G$. 
This is called the Burnside Lemma or the Burnside counting theorem. 
Useful references for the Burnside Lemma are  \cite{Cameron} \cite{wiki:burnside}, 
the former also provides other useful combinatoric background.

\section{ Review of Feynman graph combinatorics } \label{usualQFT}

There are many excellent articles and textbooks that deal with the Feynman graph expansion of perturbative
quantum field theory. Here we will simply review those aspects most relevant for our goals. For a relevant reference
that has more details see \cite{cvitan2}.

Perturbative quantum field theory expresses quantities of interest (for example, an amplitude $A$) as a power series
expansion in the small coupling $g$
\bea
A(g)=\sum_{k=0}^\infty A_v g^v
\eea
The coefficients $A_v$ are obtained by summing Feynman graphs with $v$ vertices $D_v$
\bea
  A_v = \sum_{D_v} C_{D_v} N_{D_v} F_{D_v}
\eea
$C_{D_v}$ is the symmetry factor of graph $D_v$, $N_{D_v}$ is a group theory factor coming from the
color combinatorics of global or gauge symmetry groups and $F_{D_v}$ is the result of integrating over the loop momenta in $D_v$.
For concreteness assume that $g$ is the strength of a $\phi^4$ interaction.
In this case the factor  $N_{D_v}$ is $1$. 
If we canonically quantize the theory, we expand about the free theory using Wick's Theorem. 
The Feynman graphs are used to compute the sum over all possible Wick contractions.
Different ways of doing the Wick contractions can lead to the same Feynman graph.
The symmetry factors $C_{D_v}$ keep track of this. 
Perturbation theory expands the exponential of the $g\phi^4$ interaction.
Consequently, $g^v$ come with a $1/v!$ from the Taylor series of the exponential.
Accounting for this factor, the number of Wick contractions leading to Feynman graph $D_v$ is $v! C_{D_v}$.
This observation can be used to generate an interesting sum rule for the symmetry factors. 
The sum of graphs with $v$ vertices and $E=2n$ external lines reproduces the complete set of Wick
contractions for $4v+2n$ fields. The total number of Wick contractions is equal to the number of
pairs that can be formed from the $4v+2n$ fields and is also equal to the sum over $D_v$ of $v! C_{D_v}$.
Consequently
\bea
v!\sum_{D_v}C_{D_v}=(4v+2n-1)!!
\eea
Note that for this sum rule to hold we must not subtract vacuum graphs. 
This sum rule provides a good test to check the symmetry factors. 
One way to count how many Wick contractions will produce the same graph, is to count the number of ways
of interchanging components which don't modify the graph. 
These are the automorphisms of the graph.
Indeed, the standard recipe \cite{peskin} for computing the $C_{D_k}$ starts by giving a set of rules to determine these
automorphisms. For $g\phi^4$ theory the rules are

\begin{itemize}

\item
for a closed propagator (one with both ends attached to the same vertex) there is a swap
which exchanges the endpoints.

\item
for $p$ propagators which each start at the same vertex and end at the same vertex (the starting 
and ending points might not be distinct) there are $p!$ transformations that permute the
propagators.

\item
for $n$ vertices that have the same structure there are $n!$ transformations that permute the
vertices.

\item
for $d$ identical graphs there are $d!$ transformations that permute the graphs.

\end{itemize}

Denote the automorphism group of a Feynman graph $D$ by ${\rm Aut}(D)$. 
The order $|{\rm Aut}(D)|$ is the product of the numbers obtained applying each rule above.
A Feynman graph built using $v$ vertices comes with a coefficient
\begin{equation}
   C_{D_v}={(4!)^v \over |{\rm Aut}(D)|}
\label{diagramcoefficient}
\end{equation}
Applying the rules to compute symmetry factors can be tricky. The excellent textbook\cite{peskin}
suggests that ``When in doubt, you can always determine the symmetry factor by counting equivalent
contractions''. As $v$ grows, this quickly becomes hopeless.
One of the results of this paper is give conceptually simple 
permutation group algorithms for symmetry factors (Appendix \ref{FeynGAP}).  

One basic quantity of interest is the number of graphs
\bea
  N_v=\sum_{D_v} 1 
\eea
contributing, since it is an important effect in determining the behavior of field theory at large orders in 
perturbation theory \cite{cvitan1}. For questions of this type the value of $F_{D_v}$ is unimportant (we are implicitly assuming they are
all roughly the same size) so that we may, for simplicity, work in zero dimensions which amounts to setting
$F_{D_v}=1$. 

Instead of focusing the discussion on any particular amplitude, it is useful and standard, to study the generating
functional of correlation functions
\bea
Z\left[ J\right] =\int \left[ D\phi\right] e^{iS+i\int d^4 x J\phi}
\eea
where
\bea
S=\int d^4 x \left({1\over 2}\partial_\mu\phi\partial^\mu\phi -{1\over 2}m^2\phi^2 - g\phi^4 \right)
\eea
The value of $Z$ at $J=0$ computes the sum of all vacuum graphs. Derivatives of $Z$ with respect to $J$
compute the sum of all Feynman graphs with $E$ external legs
\bea
  \left\langle 0|\phi (x_1)\phi (x_2)\cdots\phi (x_E)|0\right\rangle = {1\over i^E}\left.
    {\delta^n Z\left[ J\right]\over \delta J(x_1)\delta J(x_2)\cdots \delta J(x_E)}\right|_{J=0}
\eea
Another quantity of interest is the logarithm of $Z$, $W={\rm ln} Z$. The value of $W$ at $J=0$ computes the sum 
of all connected vacuum graphs. Derivatives of $W$ with respect to $J$, at $J=0$, compute the sum of all connected Feynman
graphs with $E$ external legs. This suggests another interesting question: how many connected Feynman graphs $N_v^{\rm conn}$ are
there? The connection between $Z$ and $W$ holds for graphs with the symmetry factor $C_{D_v}$ included. The relation between 
$N_v^{\rm conn}$ and $N_v$ (which does not include the symmetry factors) is different.

\section{ Feynman graphs in terms of the pair 
$( \Sigma_0 , \Sigma_1 ) $ }\label{FeynPair} 

\begin{figure}[ht]
\begin{center}
 \resizebox{!}{2cm}{\includegraphics{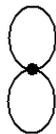}}
\caption{One vertex vacuum graph in  $\phi^4$ theory  }
 \label{fig:onevertexphi4}
\end{center}
\end{figure}

\begin{figure}[ht]
\begin{center}
 \resizebox{!}{2cm}{\includegraphics{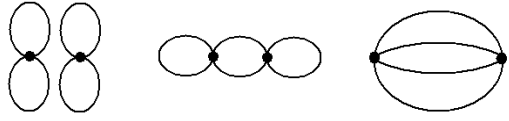}}
\caption{Two vertex vacuum graphs in  $\phi^4$ theory  }
\label{fig:twovertexphi4}
\end{center}
\end{figure}

Consider a vacuum Feynman graph in $ \phi^4 $ theory. Let $v$ be the
number of vertices.
At $v=1$ we have a single graph, given in figure \ref{fig:onevertexphi4}. 
At $v=2$ we have three graphs given in figure \ref{fig:twovertexphi4}. 
To get a systematic counting, we describe these graphs in terms of 
some numbers. One way to do this is to label the vertices $1,2, \cdots , v $, 
and list the edges as $ [ij] $. This method is well-known in graph theory 
and quickly leads to elegant results for graphs where there are no edges 
connecting a vertex to itself, and no multiple edges between a given pair of 
vertices. In the case at hand, we do have loops and we do have multiple edges between 
the same vertices. For this reason, the graphs at hand are sometimes called 
multi-graphs in the mathematics literature. 

\begin{figure}[ht]
\begin{center}
 \resizebox{!}{2cm}{\includegraphics{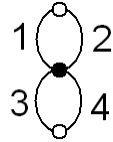}}
\caption{Numbering the half-edges}
 \label{fig:numberingHalfEdges}
\end{center}
\end{figure}

\begin{figure}[ht]
\begin{center}
 \resizebox{!}{3cm}{\includegraphics{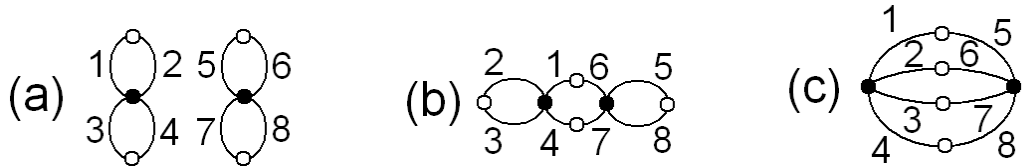}}
\caption{Numbering the half-edges  }
 \label{fig:numberingHalfEdges4vertex}
\end{center}
\end{figure}

For the graphs we consider, it is more convenient to give a combinatoric description, by 
putting a new vertex in the middle of each edge. Each edge is thus divided into two 
half-edges. Each half-edge is labeled by a number from $ 1, 2, \cdots 4v $. 
Using this labeling, the graphs shown in figure \ref{fig:onevertexphi4} 
and figure \ref{fig:twovertexphi4} have been labeled in figure \ref{fig:numberingHalfEdges} 
and figure \ref{fig:numberingHalfEdges4vertex} respectively.
We have drawn the original vertices as black dots and the new vertices as 
white dots. 

Now the graphs can be described by specifying the  list of 4-tuples of numbers at 
the black vertices and the pairs of numbers at the white vertices.  We thus introduce 
two quantities $ \Sigma_0 , \Sigma_1 $. For the $v=1$ graph, with the labeling chosen  
in figure \ref{fig:numberingHalfEdges}
\bea 
\Sigma_0  & = &  < 1 , 2, 3, 4 > \cr 
\Sigma_1 &=  &  (12) ( 34) 
\eea

For the three graphs at $v=2$, inspection of figure \ref{fig:numberingHalfEdges4vertex} shows 
that we have fixed 
\bea 
 (a) \hskip2cm \Sigma_0  &=&   < 1,2,3,4 > < 5,6,7,8 > \cr 
 \Sigma_1 & = &  (12) (34)(56)(78)   \cr 
& & \cr
 (b) \hskip2cm \Sigma_0  & = &   < 1,2,3,4 > < 5,6,7,8 > \cr 
 \Sigma_1 & = &   (23) ( 16) ( 4 7 ) ( 58  )  \cr 
& & \cr   
 (c ) \hskip2cm 
\Sigma_0  & = &   < 1,2,3,4 > < 5,6,7,8 > \cr
 \Sigma_1 & = &  (15)(26)(37)(48) 
\eea

\begin{figure}[ht]
\begin{center}
 \resizebox{!}{2cm}{\includegraphics{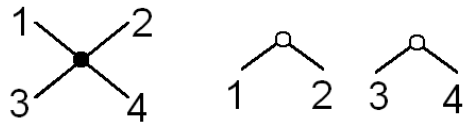}}
\caption{Splitting  the half-edges  }
 \label{fig:breakingHalfEdges}
\end{center}
\end{figure}

\begin{figure}[ht]
\begin{center}
 \resizebox{!}{4cm}{\includegraphics{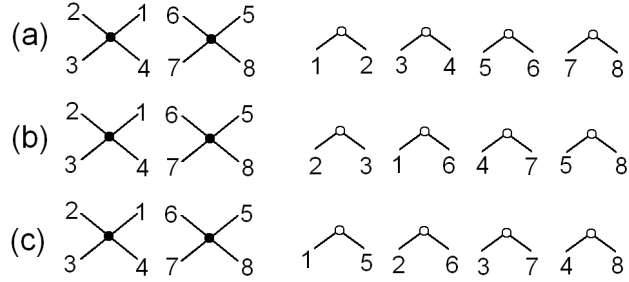}}
\caption{Splitting  the half-edges  }
 \label{fig:breakingHalfEdges4vertex}
\end{center}
\end{figure}

Clearly, we could have labeled things differently producing different pairs $ \Sigma_0 , \Sigma_1 $. 
These relabelings are elements of $S_4$ in the case $v=1$ and elements of $S_8$ in the case 
$v=2$. More generally, we have $S_{4v}$.  It is useful to consider breaking the half-edges, leaving us with 
$4v$ half-edges dangling from $v$ 4-valent vertices. And $4v$ half-edges from $2v$ 2-valent 
white vertices. We imagine we did not remove the labels as we broke the half-edges, so we know 
how to put the graph back together:  by connecting half-edges with identical numbers. 
After breaking the half edges in figures \ref{fig:numberingHalfEdges} and \ref{fig:numberingHalfEdges4vertex} we
obtain figures \ref{fig:breakingHalfEdges} and \ref{fig:breakingHalfEdges4vertex}.

The symmetry group of the 4-valent vertices includes the $4!$ permutations of 
each vertex, along with the $v!$ permutations of the vertices themselves. These permutations 
form a subgroup of $S_{4v}$ called the wreath product $S_v[S_4]$. 
It is  the semi-direct product of $(S_4)^v \rtimes S_v $. The physics reader is not required to 
have any prior knowledge of these groups.  All the relevant facts we will need from the 
math literature will be quoted as we need them. 

The symmetry group of the Feynman graph, whose order appears in the denominator 
as the symmetry factor, is obtained from the action of the symmetric group $S_{4v}$ 
on the pairs $( \Sigma_0, \Sigma_1)  $. A permutation in $S_{4v}$ is a map from integers $ \{ 1, 2, \cdots 4v \} $ 
to these integers $ \{ 1, 2, \cdots , 4v \}$, the map being one-one and onto. It can be 
written as an ordered list of the images of $\{ 1, 2, \cdots , 4v \} $, e.g a cyclic permutation 
of $ \{ 1 , 2, 3,4 \} $ can be written as 
\bea 
2341
\eea
Alternatively we use cycle notation, $(1234)$ for the example at hand. 
The image of every integer in the bracket is the one to the right, the image of the last 
is the first. 

A permutation $ \sigma $ acts on $ \Sigma_0 , \Sigma_1$ by taking the integers 
$i$ in these tuples to $ \sigma(i)$. We denote this action by 
\bea 
\sigma  : ( \Sigma_0 , \Sigma_1 )  \rightarrow ( \sigma( \Sigma_0 ) , \sigma ( \Sigma_1 ) ) 
\eea
If $ ( \Sigma_0' , \Sigma_1'  ) =  ( \sigma( \Sigma_0 ) , \sigma ( \Sigma_1 ) )  $ for some 
$ \sigma $, then both pairs $ ( \Sigma_0' , \Sigma_1'  )  $ and $ ( \Sigma_0 , \Sigma_1 ) $ 
represent the same Feynman graph. We may say that { \bf Feynman graphs correspond to orbits 
of the $S_{4v}$ action on the pairs $ ( \Sigma_0 , \Sigma_1 )$}.  

As we saw in the examples above we can fix the form of $ \Sigma_0$. 
The subgroup of $S_{4v}$ which preserves this 
is the $ S_{v} [ S_4] $. Then different Feynman graphs 
correspond to orbits of $S_{v} [ S_4 ] $
acting on the pairings $ \Sigma_1$. These are nothing 
but elements of the conjugacy class $[2^{2v}] $ in $S_{4v}$. 
This leads directly to the computation of numbers of 
Feynman graphs and symmetry factors 
 using GAP software, which is the mathematical package for 
group theory. See Appendix \ref{FeynGAP}.

\section{  $\phi^4$ theory  and string theory  } \label{sec:string} 

\begin{figure}[ht]
\begin{center}
 \resizebox{!}{4cm}{\includegraphics{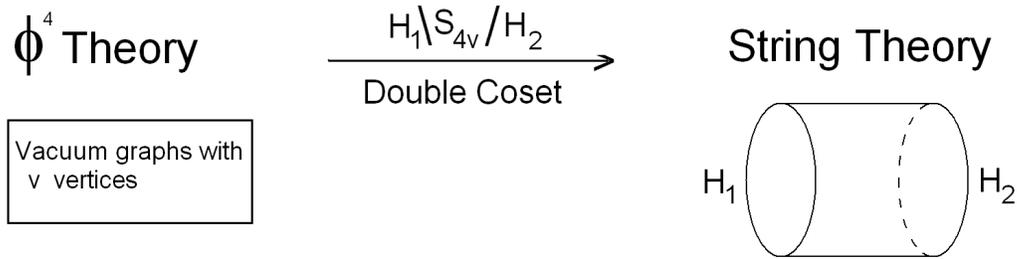}}
\caption{Double coset connection}
 \label{fig:dcoset}
\end{center}
\end{figure}

In this  section we develop a number of key ideas.
As we saw in the last section, to generate all possible Feynman graphs, we can fix $\Sigma_0$ and allow $\Sigma_1$ to vary. Using the action 
\bea 
\sigma :  ( \Sigma_0 , \Sigma_1 ) \rightarrow  ( \sigma ( \Sigma_0)  , \sigma( \Sigma_1 )  )
\eea
of $ \sigma \in S_{4v}$, the condition  $\sigma ( \Sigma_0) = \Sigma_0$
fixes $\sigma $ to be in the subgroup  $S_v[S_4]$ of $S_{4v}$.
Permutations $ \Sigma_1$ that can not be related by   $ \sigma \in S_v[S_4]$ 
are distinct Feynman graphs. This leads to the realization of
 Feynman graphs as orbits of the vertex symmetry group acting on
 Wick contractions.
A simple application of the Burnside Lemma in section \ref{sec:commperm}
then gives an explicit formula (\ref{BurnTorus})
for the number of Feynman graphs. 
The formula has an immediate interpretation as a
sum over covers of a torus. This suggests an interpretation
as string worldsheets mapping to a torus. However, the sum over maps to 
torus have some constraints which seem non-trivial to implement 
geometrically. This motivates a deeper look at the combinatorics, which reveals
a somewhat simpler connection to strings with  a cylinder target space. 
In section \ref{sec:dubcoset} we explain that Feynman graphs are elements of a double
coset. This description is already present in \cite{Read}, where 
it is derived using an operation called graph superposition. Our 
derivation does not use this operation and  is obtained
more directly using  the half-edges introduced in  Section \ref{FeynPair}. 
The  pair $ ( \Sigma_0 , \Sigma_1)$ encountered there
finds a natural interpretation in terms of the double coset. 
 This is a key result with a number of implications. It allows us
to express the counting of Feynman graphs as well as their symmetry factors 
in terms of data in $S_n$ TFT. In turn this $S_n$ TFT data is used to construct 
covers of the cylinder. 
 The figure \ref{fig:dcoset} summarizes the 
key message  of this section. This will set the stage for Section 
\ref{sec:numfeyn} where generating functions and counting sequences 
will be given.

\subsection{ Commuting permutations }\label{sec:commperm}

A vacuum Feynman graph in $\phi^4$ theory is constructed by 
starting with $v$ vertices, each having 4 edges attached, 
and connecting up the edges, as shown in Figure 11. Without any loss of generality 
label the edges of the first 
vertex $ ( 1,2,3,4)$, the edges of the second vertex as $ ( 5,6,7,8) $, etc. 
We could have chosen different numbers (from $ (1 .. 4v )$ to go with each vertex),
but we have made a choice. 
This is a choice of 
\bea 
\Sigma_0 = \langle 1,2,3,4 \rangle \langle 5,6,7,8 \rangle \cdots 
\eea 
where the angled brackets are completely symmetric and the 
different brackets can be freely interchanged. 
The symmetry of $ \Sigma_0 $ is $ S_v [ S_4 ]$.\footnote{ Some authors 
prefer the notation $S_4\wr S_v$. We follow \cite{Read} for the 
reason he gives : that the  $ S_v [ S_4 ]$ notation connects 
nicely with the substitution formula for cycle indices} 

To complete the diagram we need to give the Wick contractions.
The Wick contractions are specified by choosing an element 
$ \Sigma_1 $ in $ [ 2^{2v} ] $. For example 
\bea 
(12) (34 ) .. ( 4v-1 , 4v ) 
\eea
corresponds to the disconnected graph with a bunch of eights. 

To construct all possible Feynman graphs 
we need to allow $\Sigma_1$ to run over all possible Wick
contractions holding $\Sigma_0$ fixed to the choice we made above. 
This is easily done by acting on $\Sigma_1$ with $S_v[S_4] $.
It follows that the distinct Feynman graphs correspond to orbits of the 
vertex symmetry group $  S_v[S_4]  $  acting in the set of pairings $[2^{2v} ]$, 
which are elements of a conjugacy class in $S_{4v}$. 
If we apply the Burnside Lemma, 
we see that the number of Feynman graphs is equal to 
\bea
\hbox{ Number of Feynman Graphs } 
= { 1\over 4^v v! } \sum_{ \gamma \in H_1 } \sum_{ \sigma \in [ 2^{2v} ]  }
\delta ( \gamma \sigma \gamma^{-1} \sigma^{-1} ) 
\label{BurnTorus}
\eea 
This is a sum over certain covers of the torus
of degree $4v$. The worldsheet is also a torus.
One of the monodromies is constrained to lie 
in the subgroup $H_1 =  S_v[S_4] $, the other lies in 
the conjugacy class $C=[2^{2v}]$. Constrained sums of this 
sort where the permutations are constrained to lie in a given conjugacy class
are naturally motivated from holomorphic maps. 
Here we are constraining one permutation to lie in a subgroup 
and one to lie in a conjugacy class. Such a constraint 
is entirely possible in $S_n$ TFT, but the 
the geometrical interpretation in terms of Hurwitz space 
is, at this point,  less clear.
 Subsequent formulae we derive will show
the emergence of a cylinder rather than a torus. The formulae 
then become  more symmetric
with respect to exchange of  $H_1$ and $H_2$.

For purposes of counting applications, 
the equation (\ref{BurnTorus}) is easy to implement in GAP, 
although related techniques involving cycle indices introduced 
in Section \ref{sec:numfeyn} will be more efficient.

Apart from the number of Feynman graphs, it is also interesting to compute the
automorphism group of a given graph. This appears, for example, in the denominator
for the formula (\ref{diagramcoefficient}) for the symmetry factor. With our choice
of a fixed $\Sigma_0$, a graph is uniquely specified by $\Sigma_1$. For this
reason we will denote the automorphism group of a given graph by 
${\rm Aut}([\Sigma_1 ]_{H_1})$. The $[\Sigma_1 ]_{H_1}$ denotes 
an equivalence class under conjugation by $H_1$. The automorphism group is 
given by those elements of $H_1$ which leave $\Sigma_1$ invariant. 
The order of the automorphism group is
\bea 
|{\rm Aut} ( [ \Sigma_1 ]_{H_1})| = 
\sum_{\gamma \in H_1} \delta ( \gamma \Sigma_1 \gamma^{-1} \Sigma_1^{-1} ) 
\eea
A symmetry $ \gamma $ and a Wick contraction $\Sigma_1$ 
specify an unbranched cover of the torus, of degree $4v$ 
with $\gamma $ constrained to lie in $H_1$ and $\Sigma_1$ constrained 
to be made of 2-cycles. 

Recall from section \ref{usualQFT} that the number of Wick contractions leading to
a particular Feynman graph is
\bea\label{numwickfeyn}
v!C_{D_v}=v! {|(S_4)^v|\over |{\rm Aut} ( [ \Sigma_1 ]_{H_1})|}
\eea
Noting that ${\rm Aut}(\Sigma_0)=(S_4 )^v \rtimes S_v$ we find 
\bea
  v! {|(S_4)^v|\over |{\rm Aut} ( [ \Sigma_1 ]_{H_1})|} = { |{\rm Aut}(\Sigma_0)|\over |{\rm Aut}(D)|}
\eea
This formula has a very natural interpretation: start by making a choice of $\Sigma_0$. 
To determine a Wick contraction, we need to specify the cycle $\Sigma_1$. Any other Wick
contraction contributing to the same graph 
can be obtained by swapping vertices or swapping the half edges at a given vertex. 
The full set of these transformations is performed by ${\rm Aut}(\Sigma_0)$. If we have an 
element of ${\rm Aut}(\Sigma_0)$ that leaves $\Sigma_1$ invariant (this by definition belongs 
to ${\rm Aut}(D)$) we do not get a new Wick contraction.
Thus, the
 formula (\ref{numwickfeyn}) is an application of the Orbit-stablizer theorem
\cite{Wiki:groupaction}.

\subsection{ Double Cosets and the meaning of the pair $(\Sigma_0 , \Sigma_1)$.  } \label{sec:dubcoset}  

Read \cite{Read} derives explicit counting formulae
for graphs by developing a formulation in terms 
of double cosets. He arrives at the double cosets 
by a procedure of graph superposition. We will arrive 
at the same double coset descriptions using the 
procedure of separating  the edges into half-edges, 
and keeping track of the permutations which contain the information 
of how the half-edges are glued to make up the graph (See Figure 
\ref{fig:breakingHalfEdges4vertex}). 

There are two related double coset descriptions 
relevant to $\phi^4$ graphs. The first one 
is a double coset of $ S_n \times S_n $. The second
is a double coset of $S_n$ and can be obtained 
by a gauge fixing of the $S_n \times S_n$ picture. 
Let us start with this description. The elements of the double coset are pairs 
\bea 
 ( \sigma_1 , \sigma_2 ) \in ( S_{n} \times S_{n} ) 
\eea
with the equivalence 
\bea 
( \sigma_1 , \sigma_2 ) \sim ( \alpha \sigma_1 \beta_1 , \alpha \sigma_2 \beta_2 ) 
\eea
where $n=4v, \alpha \in S_{4v} , \beta_1 \in H_1= S_v[S_4] $ and
 $\beta_2 \in H_2= S_{2v}[S_2] $. 

To understand why this double coset counts Feynman graphs of $\phi^4$ theory consider graphs with $v$ 4-valent vertices 
which will have a $ \Sigma_0 , \Sigma_1 $ description 
\bea 
( \Sigma_0 , \Sigma_1 ) \in ( <4^v > , [ 2^{ 2v} ] )  
\eea 
As explained in section \ref{FeynPair} the graph splits up into $2v$ bi-valent white vertices 
and $v$ 4-valent black vertices. Label  the edges connected to the white vertices 
${  1 \cdots 4v } $, so we have edges $\{ W_1, W_2 \cdots W_{4v} \}$. 
Label the  edges connected to the black vertices $ \{ 1 , 2, \cdots , 4v \} $, 
so we have $ \{ B_1 , B_2 , \cdots B_{4v} \} $. 
All possible relabellings of  the edges of the white vertices are paramatrized by $ \sigma_1 \in S_{4v} $. 
All possible relabellings of the edges of the 4-valent black vertices are parametrized by $ \sigma_2 \in S_{4v} $.
Given any $ ( \sigma_1, \sigma_2)$ we can construct a graph by gluing 
\bea 
W_{ \sigma_1(i)  } \leftrightarrow B_{\sigma_2(i) } 
\eea
for $ i  = 1 \cdots 4v $.

Clearly by considering all possible $  ( \sigma_1, \sigma_2)$ we can get all possible graphs.  

If we replace the $i$ by $ \alpha(i)$, we get the same graph, since the labelings do not matter. 
So we learn that $( \alpha \sigma_1 , \alpha \sigma_2 ) $ 
and $ ( \sigma \sigma_1 , \sigma \sigma_2 ) $
produce the same graph. Hence we are interested in equivalence classes
\bea
( \sigma_1 , \sigma_2 ) \sim ( \alpha \sigma_1 , \alpha \sigma_2 ) 
\eea
  
We also know that the disconnected graph of bi-valent vertices has a symmetry 
of $ H_2 =  S_{2v}[S_2] $, so that $ \sigma_2 $ and $ \sigma_2 \beta_2 $ 
with $ \beta_2 \in H_2$ is the same graph. In the same way, $ \sigma_1 \sim \sigma_1 \beta_1 $
with $\beta_1\in H_1$.
This leads us to the conclusion that Feynman graphs are equivalence classes of the
equivalence relation
\bea\label{dcos}  
( \sigma_1  , \sigma_2) \sim ( \alpha \sigma_1 \beta_1 , \alpha \sigma_2 \beta_2 ) 
\eea
completing the demonstration that
Feynman graphs are in one-to-one correspondence with elements of a double coset.

\begin{figure}[ht]
\begin{center}
 \resizebox{!}{4cm}{\includegraphics{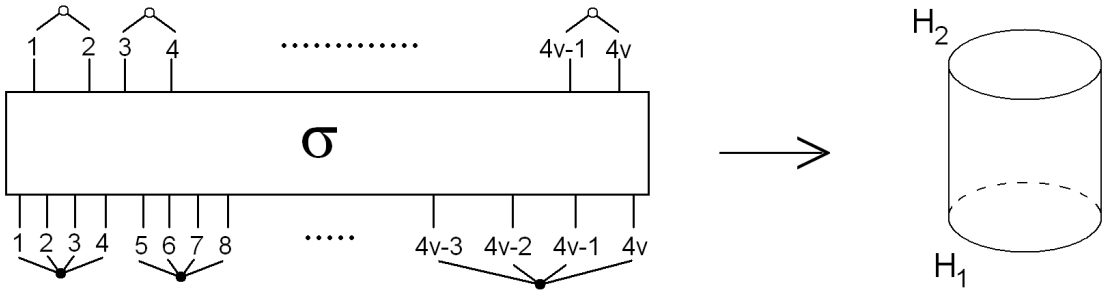}}
\caption{Double coset connection}
 \label{fig:dcoset2}
\end{center}
\end{figure}

If we choose $ \alpha = \beta_1^{-1} \sigma_1^{-1} $, then any pair is mapped to 
\bea 
( 1 , \beta_1^{-1} \sigma_1^{-1 } \sigma_2 \beta_2 ) 
\eea
So now only the combination $ \sigma_1^{-1 } \sigma_2 $ appears which is in $S_n$, 
and we have equivalences by right multiplication with $H_2$ and left multiplication 
with $H_1$. We learn that 
\bea 
 S_n \setminus ( S_n \times S_n ) / ( H_1 \times H_2) = H_1 \setminus S_n / H_2 
\eea

From (\ref{dcos}) we can consider first doing the coset by $H_1 $ and $H_2$, to we have an 
action of $S_n $ on $ S_n / H_1 \times S_n / H_2 $. This is the description we are using 
when we impose $ S_n$ relabeling equivalences on $ ( \Sigma_0, \Sigma_1 ) $ which are nothing 
but a parametrization of $ S_n / H_1 \times S_n / H_2 $.  So we have 
now learnt the group theoretic meaning of the $ \Sigma_0, \Sigma_1$, 
which we wrote earlier as a combinatoric description of the graph 
that followed (at the beginning of section 4) 
 immediately from introducing the white vertices
to separate the existing vertices and keep track of the Wick contractions.

This line of argument makes the generalization to ribbon graphs clear.
Not any permutation of the legs connecting to a black vertex is allowed: to preserve the genus
of the ribbon the cyclic order of the labeling must be respected. Consequently, $H_1= S_v[S_4]$
must be replaced by $H_1=  S_v[\mZ_4] $. 
We pursue this direction in section \ref{sec:ribbon}.

\subsection{ Cycle Index formulae related to double cosets  }\label{Nformula}

In this section we would like to make a connection with the classic results of Read \cite{Read} on the
counting of locally restricted graphs. The starting point of \cite{Read} it to count equivalence classes 
of pairs $( a_1 , a_2 ) $ in 
$(S_n \times S_n)$, where the pair $(a_1, a_2)$ is equivalent to $( b_1 , b_2 )$ if 
\bea 
( b_1 , b_2 ) = ( x a_1 g_1 ,  x a_2 g_2 ) 
\eea 
where $ x\in S_{n} $ and $ g_1 \in H_1  , g_2 \in  H_2$. This is equivalent to counting
elements of the double coset
\bea 
 S_n \setminus ( S_n \times S_n ) / (H_1 \times H_2) =  
 H_1 \setminus S_n  / H_2
\eea 
The number of equivalence classes is counted by 
$ N(Z(H_1)*Z(H_2))$, where the $Z$'s are the cycle indices, 
and $N$ is a star-product for multivariable polynomials defined as follows:
Consider two polynomials $P$ and $Q$ in variables $ f_1, f_2 .. f_n $ 
\bea 
&& P = \sum_{ j \vdash n } P_{j_1,j_2 .. , j_n } f_1^{j_1} f_2^{j_2} \cdots f_n^{j_n } \cr 
&& Q =  \sum_{ j \vdash n } Q_{j_1,j_2 .. , j_n } f_1^{j_1} f_2^{j_2} \cdots f_n^{j_n } 
\eea
where $j $ is a partition of $n$ :  $ j_1  + 2j_2 + \cdots + nj_n = n $. We 
abbreviate the coefficients as $P_j , Q_j $. The function $N$ of a 
star product is now defined by 
\bea 
N ( P * Q ) = \sum_{ j \vdash n  } P_{j}  Q_{j} {\rm Sym}( j ) 
\label{strry}
\eea
where ${\rm Sym}(j)$ is the symmetry of the conjugacy class corresponding to partition $j$
$$
  {\rm Sym}(j)=\prod_{i=1}^n (j_i)^i i!
$$ 
For our applications we will sometimes want to change the numerator group, in which case
the symmetry of the conjugacy class will change. This 
 generalization of Read's formula is discussed  in section \ref{dblcst}.

Read \cite{Read} also proves
\bea 
  N(Z(H_1)*Z(H_2))=\sum_{ p \vdash 4v } Z^{S_v[S_4] }_p Z^{ S_v [S_2] }_p Sym ( p )
                  ={1\over |H_1| |H_2| n !  } \sum_{ a_1 , a_2 \in S_{4v} }   \nu ( a_1 , a_2 ) 
\eea
where $\nu(a_1,a_2)$ is the order of the intersection $a_1 H_1 a_1^{-1}\cup a_2 H_2 a_2^{-1}$. 
We can rewrite the above as 
\bea\label{deltaRead}
N(Z(H_1)*Z(H_2)) 
&& = { 1 \over |H_1| |H_2| n !  } \sum_{ a_1 , a_2 \in S_{4v} }   \sum_{ u_1 \in H_1 } \sum_{ u_2 \in H_2 } 
\delta ( a_1 u_1 a_1^{-1} a_2 u_2^{-1} a_2^{-1})\cr 
&& = { 1 \over|H_1|  |H_2| }  \sum_{ b \in S_{4v} }  \sum_{ u_1 \in H_1} \sum_{ u_2 \in H_2}     \delta  ( u_1 b u_2^{-1} b^{-1} )  
\eea
which is reminiscent of a {\it cylinder partition function} for gauge theory with $S_{4v} $ gauge 
symmetry, and with holonomies at the boundaries restricted being summed in  $H_1 $ and $H_2$ respectively. 

Since these formulae will come up, in slight variations, repeatedly, we make some comments on notation. We will denote the number of 
points in $ H_1\setminus G/H_2$ as $\cN ( H_1 , H_2 )$ when $G$ is just the symmetric group $S_n$, or   $\cN ( H_1 , H_2 ; G )$ more generally. 
The function of $N$ of star products will also sometimes  written to make the 
numerator group explicit. So the previous formulae can be expressed as 
\bea 
\cN ( H_1 , H_2   ) = N ( Z ( H_1)  * Z ( H_2 )  )  
\eea
or 
\bea 
\cN ( H_1 , H_2   ; G = S_n ) = N ( Z ( H_1)  * Z ( H_2 ) ;  G = S_n )
\eea

Note that we can also write (\ref{deltaRead}) as  
\bea\label{charsum}  
{ 1 \over |H_1| |H_2 |  } \sum_{  R \vdash S_{4v} } \sum_{ u_1 \in H_1 } \sum_{   u_2 \in H_2 }  \chi_R ( u_1 ) \chi_R ( u_2 ) 
\eea
The sums over $u_1$ and $u_2$ produce projection operators which project onto the trivial representation so that
the only representations $R$ which contribute are the representations of $S_{4v}$ which contain the trivial of 
$S_v[S_4] $ and the trivial of $S_{2v}[S_2] $. Each such $R$ contributes the product of the multiplicities 
with which the trivial of $H_1$ and $H_2$ appear. 
\bea\label{multipls-count2}
\hbox { Number of Feynman Graphs } = \sum_{ R \vdash 4v  } \cM^{ R }_{ {\bf 1}_{H_1} }
                                                 \cM^{ R }_{ {\bf 1}_{H_2} }
\eea
We have used the notation  $\cM^{ R }_{{\bf 1}_{H_1}}$ for the multiplicity of the one-dimensional representation 
of $H_1$ when the irreducible representation $R$ of $S_{4v} $ is decomposed into representations of  the subgroup $H_1$. 


\subsection{ Action of 
Vertex symmetry group on Wick contractions }\label{vertexonwick}

We now have two different ways to compute the number of Feynman graphs: as the number of orbits
of the vertex symmetry group acting on Wick contractions, which leads to
\bea\label{deltaUS} 
{ 1 \over (4!)^v  v! } \sum_{ u_1   \in  S_v [S_4] } \sum_{ \sigma \in [ 2^{2v} ] } \delta (  u_1 \sigma u_1^{-1} \sigma^{-1} ) 
\eea
or in terms of  the star product of the cycle indices $N(Z(H_1)*Z(H_2))$. We will demonstrate, using some general group theory,
the equivalence of (\ref{deltaUS})and (\ref{deltaRead}). 

Noting that $ H_1 = S_v[S_4]$, (\ref{deltaUS}) and (\ref{deltaRead}) are equivalent if
\bea\label{equalityconjgp} 
\sum_{ \sigma \in [ 2^{2v} ] } \delta ( u_1 \sigma u_1^{-1} \sigma^{-1} ) 
= { 1 \over | H_2 | } \sum_{ b \in S_{4v} } \sum_{ u_2 \in H_2 } \delta ( u_1 b u_2^{-1}  b^{-1} )  
\eea
To see that this relation holds, suppose there is  a solution to the first delta function. 
$\sigma $ is conjugate by some $b$ to $\sigma_0\equiv (12)(34) .. (4v-1,4v )$
\bea 
b^{-1}  \sigma_0 b = \sigma
\eea
We know that all elements commuting with $\sigma_0$ are in $H_2 \equiv 
 S_{2v}[S_2] $. 
From (\ref{equalityconjgp}) we have 
\bea 
u_1 \sigma u_1^{-1} = \sigma 
\eea
So 
\bea 
u_1 b^{-1}  \sigma_0 b  u_1^{-1} = b^{-1} \sigma_0 b  
\eea
It follows that 
\bea 
b u_1 b^{-1}  = u_2 
\eea
for some $u_2$ in $H_2$. 
 The $b$ that takes $\sigma_0$ to $\sigma$ can be multiplied by an arbitrary element of $H_2$. 
So if we replace the sum over $\sigma\in [2^{2v}]$ by a sum over $b, u_2$, we will be over counting by 
$|H_2|$. So we conclude that the equality (\ref{equalityconjgp}) is correct.

It is instructive to note that the set of elements $[2^{2v}]$ can be thought as the set
of cosets $S_{4v}/H_2$, where $H_2= S_{2v}[S_2] $. To see this, start by noting that
$S_{4v}$ acts on $\sigma_0$ by conjugation to generate all the elements 
in $[2^{2v}]$. For any $\sigma\in [2^{2v}]$, we have 
\bea 
\sigma = \alpha \sigma_0 \alpha^{-1} 
\eea
for some $ \alpha \in S_{4v} $.  If we multiply $ \alpha $ on the right 
by $ \beta \in S_{2v} [S_2]  $, we get the same $\sigma $. We can write 
\bea 
[2^{2v}] = S_{4v}/H_2 
\eea
Thus, we can rewrite (\ref{deltaUS}) as 
\bea\label{deltaUScoset}  
\hbox{ Number of Feynman Graphs } = 
{ 1 \over (4!)^v  v! } \sum_{ u_1   \in H_1  }
 \sum_{ \sigma \in S_{4v}/H_2 } \delta (  u_1 \sigma u_1^{-1} \sigma^{-1} ) 
\eea

A similar argument implies that 
\bea 
< 4^v > = S_{4v}/ H_1 
\eea
Here $ < 4^v> $ stands for the space of $ \Sigma_0 $'s. 

Using that fact that the expression (\ref{deltaRead}) is symmetric under the exchange of 
$H_1 $ and $H_2$, we can also write 
\bea\label{deltaUScoset2}  
\hbox{ Number of Feynman Graphs } 
=  { 1 \over 2^v (2v)!  } \sum_{ u_2   \in H_2  }
 \sum_{ \Sigma_0 \in S_{4v}/H_1 } \delta (  u_2 ( \Sigma_0)  ,  \Sigma_0 )
\eea
Thus, by Burnside's Lemma the number of Feynman graphs is also equal to the number of orbits of
$ S_{2v} [S_2] $ acting on the set of vertex labels.
Note that (\ref{deltaUScoset2}) looks slightly 
different from (\ref{deltaUScoset})
because $\Sigma_1$ is a permutation, but $\Sigma_0$ is not. 
The action of substituting $i \rightarrow u(i)$ for some permutation
 $u \in S_{4v} $ can be achieved by conjugation
 for $\Sigma_1$ but not for $\Sigma_0$.

\subsection{ Number of Feynman graphs from strings on a cylinder 
} \label{relate2string}

In this subsection we show that the formula (\ref{charsum}) for 
counting Feynman graphs in $\phi^4$ theory is computing 
an observable in $S_n$ TFT. 
 Recall from section \ref{2dYM} that
the expectation value of the observables $tr (\sigma_1 U_1^{\dagger}), tr (\sigma_2 U_2^{\dagger})$
on the cylinder with boundary holonomies $U_1$ and $U_2$ are
\bea 
Z ( \sigma_1 , \sigma_2 ) = \int dU_1 dU_2 tr ( \sigma_1 U_1^{\dagger}  ) tr ( \sigma_2 U_2^{\dagger} ) 
  Z ( U_1 , U_2) 
=\sum_{ \gamma } \delta ( \sigma_1 \gamma \sigma_2 \gamma^{-1} ) 
\label{deltsum}
\eea
Summing this expectation value over permutations in the subgroups $H_1, H_2$ 
\bea\label{sumsH1H2}  
Z ( H_1 , H_2 ) && \equiv { 1 \over | H_1|  | H_2 |  }\sum_{ \mu_1 \in H_1  } \sum_{ \mu_2 \in H_2 } Z ( \mu_1 , \mu_2 ) \cr
 && =  { 1 \over | H_1|  | H_2 |  }\sum_{ \mu_1 \in H_1  }
 \sum_{ \mu_2 \in H_2 }  \sum_{ \sigma \in S_n }
  \delta ( \mu_1 \sigma \mu_2 \sigma^{-1} )  
\eea
we recover the formula (\ref{deltaRead}) for counting Feynman graphs in $\phi^4$ theory.

There is an interesting subtlety we should comment on. 
In the large $N$ expansion of $U(N)$ 2dYM, we encounter 
observables which can be parametrized using permutations. 
For observables constructed from gauge-invariant polynomials 
of degree $n$ in the holonomy $U$, the partition functions 
are functions only of the conjugacy class of $\sigma $ in $S_n$.  
In the above expression (\ref{sumsH1H2}), the sums over $\mu_i$ 
do not run over entire $S_n$ conjugacy classes, but are restricted to $H_i$.  

As explained in section 2.2 the observables in 
the $\Omega \rightarrow 1$ approximation of large $N$ 2dYM, 
for Riemann surfaces with or without  boundary, 
can be expressed in terms of physical observables 
in $S_n$ TFT. The quantity in 
(\ref{sumsH1H2}) is a generalization of the observables one gets
from the large $N$ expansions of $U(N)$ 2dYM, but it is still 
an observable in the lattice $S_n$ TFT. The boundary observables 
are not invariant under conjugation by $S_n$ elements at the 
boundary. The $S_n$ conjugation symmetry is broken to $H_1$ and $H_2$ 
respectively. 
There is a general principle of Schur-Weyl duality which relates the
symmetric group constructions to unitary groups 
(see \cite{CMR,tenyears,ehs,robert} for applications in
gauge-string duality). The wreath products of symmetric groups have also
appeared in 
connection with symmetrised traces in the context of constructing
eighth-BPS operators \cite{countconst}. We therefore expect that these more 
general $S_n$ TFT observables can also be expressed in terms of some 
construction with gauge theory involving unitary groups. We
will leave this clarification for the future.

As mentioned in Section 2, the $S_n$  TFT  
is closely related to covering space theory.
In the next section we will encounter 
expressions of the form (\ref{sumsH1H2}) 
but without the sum over $\sigma$. We will 
show how cutting and gluing constructions
of covering spaces use the data $ \mu_1 , \mu_2 , \sigma $.

\subsection{ The symmetry factor of a Feynman graph from strings on a cylinder   }
\label{sec:symmfacstrings}

\begin{figure}[ht]
\begin{center}
\resizebox{!}{4cm}{\includegraphics{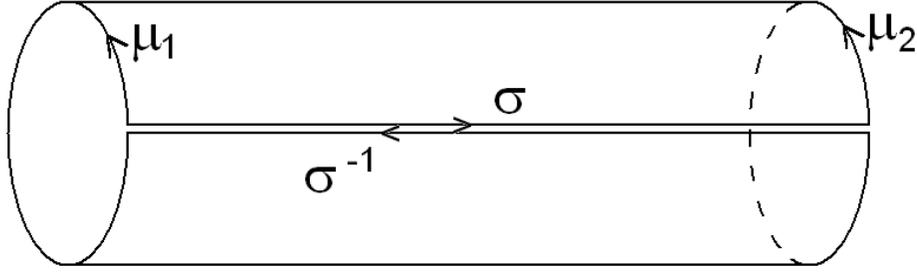}}
\caption{ The symmetry factor of a graph is an open Wilson line $S_n$ lattice TFT observable  }
\label{fig:cylindersymm}
\end{center}
\end{figure}

We have already argued that the Feynman graphs of $\phi^4$ theory correspond to 
elements of the double coset, 
\bea\label{thephi4dubcos}  
    S_{4v}     \setminus  ( S_{4v} \times S_{4v} )     / 
( ~ S_{v} [ S_4] \times S_{2v} [S_2] ~ )
\eea
The $S_{4v}$ on the left is the diagonal subgroup of the product group, i.e 
pairs of the form $ (\sigma , \sigma)$. 
The symmetry factor of a Feynman graph corresponding to an orbit with 
representative
$(\sigma_1 , \sigma_2)$ is the size of the 
stabilizer group. This can be computed by calculating 
\bea 
\sum_{ \gamma \in S_{4v} } \sum_{ \mu_1 \in   S_{v} [ S_4 ]  } 
\sum_{ \mu_2 \in  S_{2v} [S_2] } 
\delta ( \gamma \sigma_1 \mu_1 \sigma_1^{-1}  ) 
\delta ( \gamma \sigma_2 \mu_2  \sigma_2^{-1}  )
\eea
Using one of the delta functions to perform the sum over $\gamma$ we obtain
\bea 
 \sum_{ \mu_1 \in   S_{v} [ S_4 ]  } 
\sum_{ \mu_2 \in  S_{2v} [S_2] } 
\delta (  \sigma_1 \mu_1^{-1}  \sigma_1^{-1}  \sigma_2 \mu_2 \sigma_2^{-1}   )
\eea
By defining $\sigma = \sigma_1^{-1}\sigma_2$, 
we can write the formula for the symmetry
factor as 
\bea\label{symmfactorAmp}  
{\rm Sym} ( \sigma )  =  \sum_{ \mu_1 \in   S_{v} [ S_4 ]  } 
\sum_{ \mu_2 \in  S_{2v} [S_2] } 
\delta (  \mu_1^{-1}  \sigma \mu_2 \sigma^{-1}    )
\eea
This expression relates more directly to the equivalent description of the double coset (\ref{thephi4dubcos}) as 
\bea 
     S_{v} [ S_4]    \setminus   S_{4v}      / 
\times S_{2v} [S_2] 
\eea
Comparing to (\ref{deltasum}) we see that this is
an  observable  in $S_n$ TFT, 
 with $H_1, H_2$ observables on the boundaries. This is illustrated in 
Fig. \ref{fig:cylindersymm}. 

Consider a cover of the cylinder. Choose a point on one boundary circle
and label the inverse images of that point $ \{ 1 , 2, \cdots , n \} $. 
Following the inverse images of these points along the boundary circle
leads to a permutation $ \mu_1$ which is constrained to be in $H_1$. 
Likewise there is a permutation $\mu_2$  of $ \{ 1' , 2' , \cdots , n' \} $
at the other boundary. Following a path on the cylinder 
which joins the two points will produce a permutation $\sigma$. 
We are fixing $ \sigma $ to lie in a fixed element of 
$H_1 \setminus  S_n/H_2$. 

From the point of view of $S_n$ lattice TFT,  
the line joining the two chosen points on the boundary 
circle, associated with a fixed permutation $ \sigma $ 
is a Wilson line.
Given the close relation, between  $S_n$ lattice TFT
and Hurwitz spaces, we can map the Wilson line observable 
to a construction in covering spaces of the cylinder.

\begin{figure}[ht]
\begin{center}
\resizebox{!}{4cm}{\includegraphics{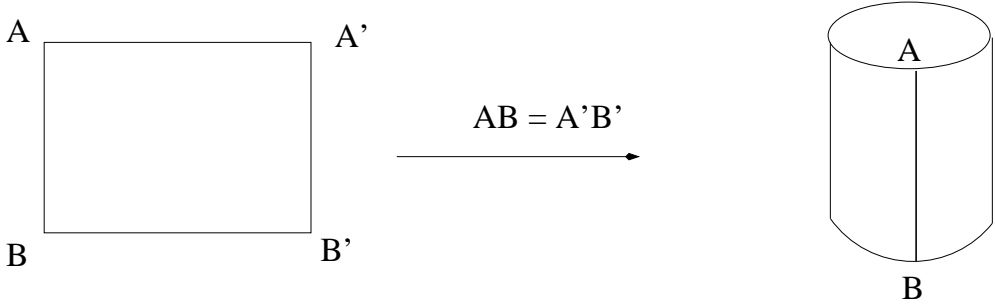}}
\caption{ Gluing a square to get cylinder }
\label{fig:gluesquare}
\end{center}
\end{figure}

\begin{figure}[ht]
\begin{center}
\resizebox{!}{4cm}{\includegraphics{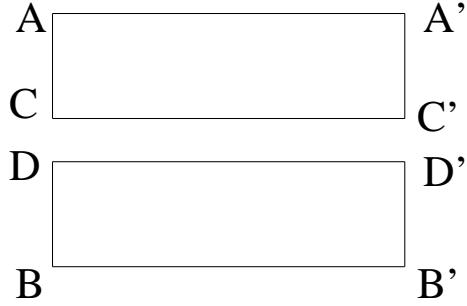}}
\caption{ Gluing a pair of rectangles to get cylinder }
\label{fig:gluerectangles}
\end{center}
\end{figure}

\begin{figure}[ht]
\begin{center}
\resizebox{!}{12cm}{\includegraphics{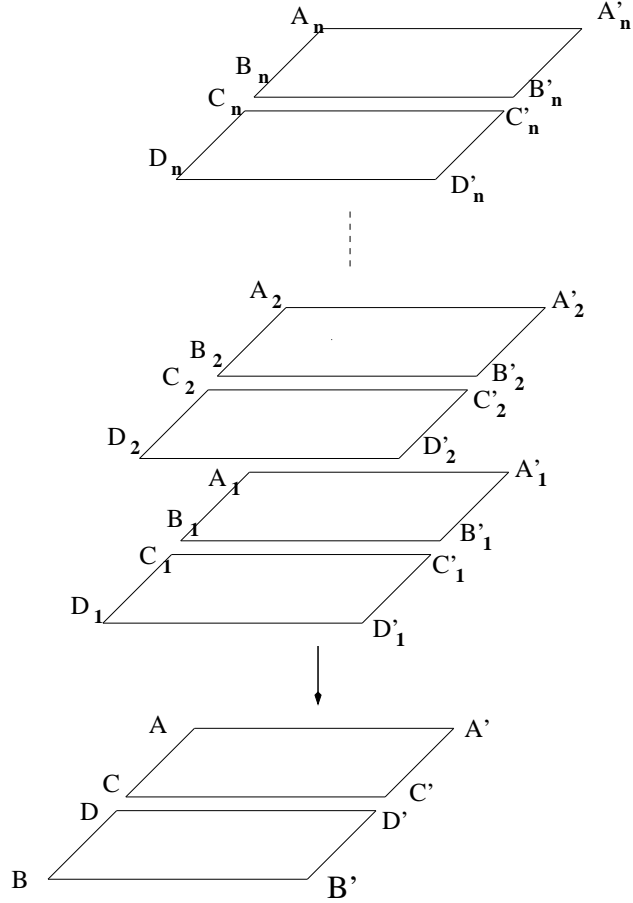}}
\caption{ Gluing copies of rectangle pairs  to get covering of cylinder - identifications determined by $\mu_1, \mu_2 , \sigma$  }
\label{fig:gluerectanglescover}
\end{center}
\end{figure}

First notice that the cylinder can be obtained
by gluing two ends of a square  $[AA'B'B]$ in Figure 
\ref{fig:gluesquare}.   This 
is a topological quotient which identifies 
the edge $AB$ with $ A'B'$. 
 To construct a covering 
space associated with the data $ \mu_1 , \sigma , \mu_2 $ 
subject to $\mu_1^{-1} \sigma  \mu_2 \sigma^{-1}  = 1 $, 
we consider cutting the square further into the rectangles 
$ [AA'C'C] $ and $[DD'B'B]$ (Figure \ref{fig:gluerectangles}). 
The cylinder is recovered by the gluings (Fig 14)
\bea\label{gluerectangles} 
&&  AC = A'C' \cr
&&  DB = D'B' \cr 
&&  CC' = DD' 
\eea
Now take $n$ copies of these pairs of rectangles, 
with labels $  [A_iA'_iC'_iC_i] $ and $[D_iD'_iB'_iB_i]$, 
with $i$ ranging from $1$ to $n$, as in Figure \ref{fig:gluerectanglescover}. 
For the application to (\ref{sumsH1H2}) we  have $ n = 4v$. 

The  union of 
 rectangles  $  \{ [A_1A'_1C'_1C_1],    \cdots ,  [A_nA'_nC'_nC_n] \} $
 is quotiented by identifying edges $A_iC_i$
to the edges  $A'_i C'_i $ by the permutation $ \mu_1$. 
\bea 
 A_i' C_i' = A_{ \mu_1 (i) }  C_{ \mu_1(i) } \cr  
 \eea
The rectangles  $[D_iD'_iB'_iB_i]$ are quotiented by 
identifications 
\bea 
D'_i B'_i = D_{\mu_2  ( i)  } B_{ \mu_2  ( i)   }  
\eea
Finally we identify 
\bea 
D_i D'_i = C_{ \sigma (i) } C'_{ \sigma(i)}  
\eea
The covering map is specified by mapping each of the
$n$ rectangles, to the original rectangles without labels
in the obvious way (see Figure \ref{fig:gluerectanglescover})

\begin{figure}[ht]
\begin{center}
\resizebox{!}{4cm}{\includegraphics{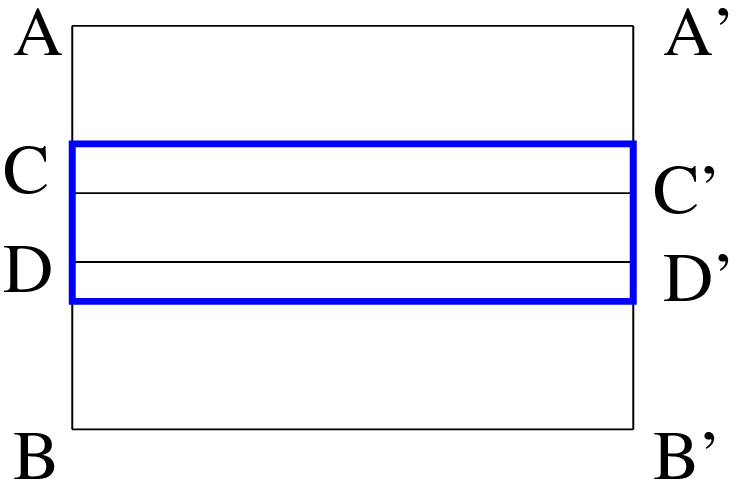}}
\caption{Closed contractible path  }
\label{fig:closedpath}
\end{center}
\end{figure}

The condition  $\mu_1^{-1}  \sigma  \mu_2  \sigma^{-1}  = 1 $
ensures that if we consider the inverse image of  
the closed contractible  path shown in blue in  Figure \ref{fig:closedpath}, 
the inverse image is also contractible.  

This type of construction is  a standard part of covering space theory
in connection with the Riemann existence theorem
(see for example \cite{ezell}).

The  generalizations of (\ref{Zsig1sig2}) visible in (\ref{sumsH1H2}) and 
(\ref{symmfactorAmp}) involve restricting the sums over 
$\mu_1, \mu_2 $ to specific subgroups $H_1, H_2$ and leaving the 
$\sigma $ unsummed. From the point of view of $S_n$  TFT 
these constraints are easily implemented, since the degrees
of freedom on each edge form the whole group, where elements 
and subgroups can be chosen. However the boundary observables 
are not invariant under conjugation by $S_n$ and have no dual 
in the large $N$ 2dYM that we yet know how to construct. 
Given the close connection between 
$S_n$ elements and monodromies of covers, it is not 
surprising that a definite covering space construction for specific 
$\mu_1 , \mu_2 , \sigma$ exists. We have provided
such an explicit cutting-and-gluing construction here. 
There are clear analogies between the construction given  and 
the implementation of twisted boundary conditions in 
D-branes on orbifolds \cite{douglasmoore}. The presence of 
a Wilson line defect located at a line joining the 2 boundaries of 
the cylinder in the $S_n$ TFT picture  suggests that  
there should be a D-brane interpretation. We leave this as a  
future direction of research.  Developing  the 
connection of the lattice $S_n $ TFT 
to the axiomatic approach to TFT  and branes in  \cite{moore-segal} maybe 
a useful approach.

In the standard discussions of the Riemann existence theorem 
for (branched) covers, one has a two-way relation. On the geometrical 
side, there are maps $f$ with the equivalence $ f = f \circ \phi$
for holomorphic automorphisms $\phi$ 
(see more details in \cite{CMR,lanzvon}). On the combinatoric side 
there are permutations with conjugation equivalence.
From permutations we can construct covers, and vice versa.
The equivalence classes map to each other. Here we have 
permutations, but with some restrictions having to do with choices
$H_1,H_2$. We can still construct some covers with this data. 
We have not fully articulated the correct equivalences on the 
geometrical side for a 1-1 correspondence. The construction we gave should 
provide useful hints for the precise definitions on the geometrical side, which 
will provide the equivalence. We leave this as a problem for the future.

\section{ The number of Feynman graphs }\label{sec:numfeyn} 

In section \ref{Nformula} a formula expressing the number of Feynman graphs in terms of the star products
of two cycle indices was given. This result is useful because formulas for the cycle index of the wreath 
product of two symmetric groups are known. Using these results we will write down rather explicit formulas
for the number of vacuum graphs in $\phi^4$ theory in section \ref{phi4vac}.

\subsection{ Some analytic expressions exploiting generating functions }\label{phi4vac}

Let $d_i$ denote the total number of vacuum Feynman graphs in $\phi^4$ theory 
with $i$ vertices. To obtain an analytic formula for $d_i$
we will need the cycle indices of two wreath products, $S_n [S_4]$ and $S_n [S_2]$.
The known generating functions for these wreath products are 
\bea 
\cZ^{ S_{\infty} [ S_4 ] } [ t , \vec x ] \equiv
\sum_{n} t^n Z^{S_n[S_4]}(\vec x ) 
= e^{ \sum_{i=1}^{ \infty}  { t^i \over 24i}  ( x_i^4 + 8   x_{3i} x_i +  6 x_{2i} x_i^2  + 3x_{2i}^2 + 6 x_{4i})}  
\eea
\bea
\cZ^{ S_{\infty} [ S_2 ] } [ t , \vec x ] \equiv
\sum_{n} t^n Z^{S_n[S_2]}(\vec x ) 
= e^{ \sum_{i=1}^{ \infty}  { t^i \over 2i}  ( x_i^2  +  x_{2i})}
\eea 

To compute $N(Z(H_1)*Z(H_2))$ take the product of the coefficients of $\prod_{i} x_i^{p_i}$ 
in the two cycle indices, and then further take the product with $\prod_i i^{p_i  } p_i ! $.
This can be accomplished if we make the substitution $ x_i \rightarrow \sqrt{i} y_i$, to obtain
\bea 
\tilde Z^{ S_{\infty} [ S_2] } [ t , y_i ] && = \cZ^{ S_{\infty} [ S_2 ] }[t ,x_i = \sqrt{i} y_i  ] \cr 
\tilde Z^{ S_{\infty} [ S_4 ] } [ t , y_i ]&& = \cZ^{ S_{\infty} [ S_4 ] }[t ,x_i = \sqrt{i} y_i  ] 
\eea
and then replace the $(y_i \bar y_i)^{p_i}$ with $p_i !$.
This leads to the following formula for the number of $\phi^4$ vacuum graphs with $v$ vertices 
\bea 
d_v= \oint {  t_1^{ -2v } dt_1 \over t_1  }\oint {  t_2^{ - v } dt_2  \over t_2  }
 \prod_{i=1}^{\infty} \int {dy_i d \bar y_i\over 2\pi} e^{ -\sum_k y_k \bar y_k } 
 \tilde Z^{ S_{\infty} [ S_2] } [ t_1  , y_i ]  \tilde Z^{ S_{\infty} [ S_4 ] } [ t_2   , \bar y_i ] 
\eea 
The integrals over $y_i$ and $\bar{y}_i$ ensure that 
only terms with equal powers of $y_i$ and  $\bar{y}_i$ contribute, and they
implement the substitution $( y_i \bar y_i)^{p_i}\to p_i !$.
The contours of integration over $t_1$ and $t_2$ are both counter clockwise and they both
encircle the origin.

We could contemplate many ways of refining this result. For example, can we determine how many of these vacuum
graphs are connected? Before turning to this question, it is instructive to ask how we can identify 
if a graph is connected or disconnected from its description in terms of the pair $\Sigma_0,\Sigma_1$.
A Feynman graph ($\Sigma_0,\Sigma_1$) with $v$ vertices is disconnected if for some subgroup $G=S_{4v-p}\times S_p$ 
with $p>0$ we have $\sigma (\Sigma_1)\in G$ for all $\sigma\in {\rm Aut}(\Sigma_0)$. Let $c_i$
denote the number of connected Feynman graphs with $i$ vertices. The total number of vacuum graphs with $i$ 
vertices is the coefficient of $g^i$ in the partition function
\bea
  Z\equiv 1+\sum_{i=1}^\infty d_i g^i=\prod_{i=1}^\infty \left[ {1\over 1-g^i}\right]^{c_i}
  \label{frstconnected}
\eea
This formula, given for example in \cite{Read}, 
can be used to determine the $c_i$ once the $d_i$ have been computed.
It is now straight forward to determine the number of vacuum diagrams. See Appendix \ref{data} for
numerical results.

\subsection{ Scalar fields with external legs } 

The next natural generalization is to consider Feynman graphs with $E$ external legs.
Summing Feynman graphs with $E$ external legs produces an $E$-point correlation function.
For the generic case when all $E$-points corresponds to different spacetime events, graphs
obtained by permuting labels of the external points give distinct contributions to the
correlation function. For this reason the automorphism group of the graph does not include 
any elements that permute the labels of external legs.

For a graph with $v$ vertices and $E$ external legs, we obtain a total of $4v+E$ half edges.
$\Sigma_1$ is now an element of the conjugacy class $[2^{2v+{1\over 2}E}]$ of $S_{4v+E}$, while
$\Sigma_0$ contains $v$ 4-tuples of numbers specifying how the half edges connecting to the vertices
are labeled, as well as $E$ numbers specifying how the half edges connecting to external points
are labeled. Since we do not have automorphisms that permute the half edges connecting to external 
points, $H_1={\rm Aut}(\Sigma_0)=S_v[S_4]\times (S_1)^E$. 
We can also write the subgroup as $S_v[S_4]$, with the understanding 
that it is acts by keeping fixed the last $E$ integers from the 
set $\{ 1 , 2 , \cdots , 4v+E \}$ that $S_{4v+E}$ acts on.  
Consequently, Feynman graphs in $\phi^4$ theory
with $v$ vertices and $E$ external points are in one-to-one correspondence with elements of the
double coset
\bea 
(S_{v}[S_4]\times S_1^{E})  \setminus  S_{4v+E}/S_{(4v+E)\over 2}[ S_2 ] 
\eea 
To compute the number of graphs the only new cycle index we need is
\bea 
 Z^{ S_{ v } [ S_4 ] \times S_1^E } ( x_1 , \cdots , x_n ) 
 = x_1^E  Z^{ S_{ v } [ S_4 ] } (  x_1 , \cdots , x_n )
\eea
It is now straight forward to obtain the number of Feynman graphs in $\phi^4$ theory with $E$
external lines
\bea 
\cN ( H_1 , H_2 ; G ) = N( Z(H_1) * Z (H_2) ;S_{4v+E}) =
 \sum_{ p \vdash ( 4v + E ) } Z_{p}^{ S_{ n } [ S_4 ] \times S_1^E } Z^{S_{  2 v +
{  E  \over 2 } }    [ S_2 ]  }_{p} {\rm Sym}( p )  
\label{Eistwo}
\eea
The notation $  p \vdash ( 4v + E ) $ indicates that $p$ runs over all 
partitions of $4v+E$. 

We can again refine this counting by asking how many of these graphs are connected. For concreteness, focus on
the case $E=2$. Let $d_{2,i}$ denote the number of Feynman graphs with two external lines and $i$ vertices and let $c_{2,i}$
denote the number of these graphs that are connected. We can again write down a partition function which gives
the total number of graphs with two external lines and $i$ vertices as the coefficient of $g^i$. In this formula
the number of vacuum graphs $c_i$ and $d_i$ defined in the last subsection participate. We find
\bea
 \prod_{i=1}^\infty \left[ {1\over 1-g^i}\right]^{c_i}\left(\sum_{k=0}^\infty c_{2,k} g^k\right) 
=\left(\sum_{k=1}^\infty d_{2,i} g^i\right)
\label{scndconected}
\eea
This formula can be used to determine the $c_{2,i}$ once the $d_{2,i}$ have been computed using (\ref{Eistwo}).
See Appendix \ref{data} for numerical results.

\subsection{ Generalization to $\phi^3$ theory and other interactions }

Although our discussion has focused on $\phi^4$ theory it should be clear that our methods are general. To illustrate
this we will now consider a $\phi^3$ interaction. A graph with $v$ vertices has $3v$ half edges.
$\Sigma_1$ is now an element of the conjugacy class $[2^{3v\over 2}]$ of $S_{3v}$, while
$\Sigma_0$ contains $v$ 3-tuples of numbers specifying how the half edges connecting to the vertices
are labeled. We have $H_1=S_v[S_3]$ and $H_2=S_{3v\over 2}[S_2]$. Vacuum Feynman graphs in $\phi^3$ theory
with $v$ vertices are in one-to-one correspondence with elements of the double coset
\bea 
S_{v}[S_3]  \setminus  S_{3v}/S_{3v\over 2}[ S_2 ] 
\eea 
To compute the number of graphs the only new cycle index we need can be read from the generating function
\bea 
 Z^{ S_{\infty} [ S_3 ] } [ t  ; x_1 , x_2 , \cdots ] 
&& = \sum_{ n=0}^{\infty} t^n Z^{ S_{ n } [ S_3 ] } ( x_1 , x_2 , \cdots , x_n ) 
\cr 
&& = e^{ \sum_{ i=1}^{ \infty } { t^i \over 6 i } ( x_i^3 + 3 x_{2i } x_i + 2 x_{3i } )  }
\eea  
It is now straight forward to obtain the number of vacuum Feynman graphs in $\phi^3$
\bea 
N(Z(H_1)*Z(H_2)) = \sum_{ p \vdash 3v  } Z_{p}^{ S_{ v } [ S_3 ]} Z^{S_{{3v \over 2}}[S_2]}_{p}{\rm Sym} ( p )  
\label{vacuumphi3}
\eea

Feynman graphs with $E$ external legs are in one-to-one correspondence 
with elements of the double coset 
\bea 
(   S_{ v} [ S_3 ] \times S_1^{ E } )  \setminus   S_{3v + E } / S_{ ( 3v + E)  \over 2 }   [ S_2 ] 
\eea
Using the generating function  
\bea 
 Z^{ S_{ n } [ S_3 ] \times S_1^E } ( x_1 , \cdots , x_n ) 
 = x_1^E  Z^{ S_{ n } [ S_3 ] } ( x_1 , \cdots , x_n ) 
\eea
it is straight forward to compute the number of Feynman graphs with $v$ vertices and $E$ edges
\bea 
\cN ( H_1 , H_2 ; S_{3v+E}) ) 
= N ( Z(H_1)  * Z(H_2) ;S_{3v+E}) = \sum_{ p \vdash ( 3v + E ) } Z_{p}^{ S_{ v } [ S_3 ] \times S_1^E } Z^{S_{ 3 v +
  E  \over 2 }     [ S_2 ] }_{p} {\rm Sym} ( p )  \cr 
\label{extlegsphi3}
\eea
The notation $  p \vdash ( 3v + E ) $ indicates that $p$ runs over partitions 
of $(3v+E)$. 
See Appendix \ref{data} for numerical results.
Cubic graphs have recently played a role in 
studies of $N=8$ SUGRA \cite{bern} as well as in the classification of 
$N=2$ 4-dimensional gauge theories \cite{nopphan}. The $\phi^3$ theory 
in 6 dimensions is known to be asymptotically free \cite{srednicki}. 
Hence this case is of special interest.

Finally, consider a theory with both cubic and quartic vertices. 
To count the vacuum graphs having $v_3$ cubic and $v_4$ quartic vertices, we would repeat the above discussion with 
$ S_{v_3} [ S_3 ] \times S_{v_4} [ S_4 ]$ replacing the groups  $S_{v_3} [ S_3 ]$ (for $\phi^3$)
or $  S_{v_4} [ S_4 ]$ (for $\phi^4$).

\subsection{ Interpretation in terms of 2dYM string }

We have already argued that the number of vacuum graphs in $\phi^4$ theory is computing an observable in the
Lattice TFT with $S_n$ gauge group.
The counting of Feynman graphs in this section also has a formulation in Lattice TFT: One
simply uses (\ref{sumsH1H2}) with the appropriate $H_1,H_2$ described above. 

The formula (\ref{symmfactorAmp}) for the symmetry factor as an amplitude is also directly applicable here.

\section{ Feynman graphs as orbits: a view of ribbon graphs }
\label{sec:ribbon}

The identification of Feynman graphs as elements of the double coset 
$S_n \setminus ( S_n \times S_n ) / ( H_1 \times H_2)$ is a nice unifying 
picture. The groups $H_1$ and $H_2$ are the symmetries of the interaction (black) 
and bivalent (white) vertices respectively. This allows a simple generalization
from ordinary graphs to ribbon graphs: symmetries must by definition preserve
the genus of the ribbon graph. The cyclic order of the labels at a black vertex
must be respected if the genus is to be preserved. Thus, for example, the 
replacement $H_1=S_v[S_4] \rightarrow H_1=S_v[\mZ_4]$ takes us from counting Feynman 
graphs in $\phi^4$ theory to counting ribbon graphs in ${\rm Tr} \phi^4$ theory.

\begin{figure}[ht]
\begin{center}
 \resizebox{!}{4cm}{\includegraphics{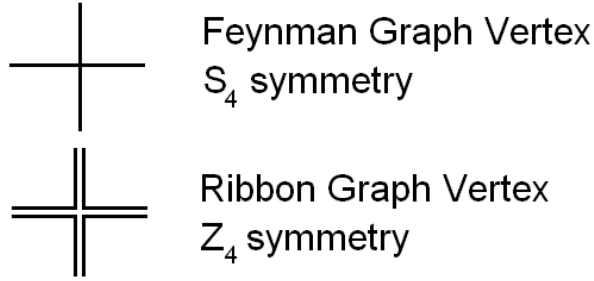}}
\caption{Cyclic versus symmetric for Feynman vertex versus ribbon vertex   }
 \label{fig:graphvertex}
\end{center}
\end{figure}

It is now clear that ribbon graphs can also identified with elements of a
 double coset. For example,
the vacuum graphs with $v$ vertices of a matrix model with ${\rm Tr}\phi^4$ interaction
are elements of the coset 
\bea 
S_{4v} \setminus  ( S_{4v} \times S_{4v} ) / ( S_v [ \mZ_4 ] \times S_{2v} [ S_2 ] ) 
\eea
where we have identified $H_1=S_v [ \mZ_4 ]$ and $H_2=S_{2v} [ S_2 ]$.


\subsection{ Number of ribbon graphs using commuting pairs } 

Using the insights of the previous subsection and repeating the argument of section \ref{sec:string},
we can generate all possible ribbon graphs by allowing $S_v[\mZ_{4}]$ to act on the set of all possible
Wick contractions, that is, the conjugacy class $[2^{2v}]$ of $S_{4v}$. Each orbit of $S_v[\mZ_{4}]$ is 
a distinct ribbon graph. Thus, the number of ribbon graphs can be obtained, using Burnside's Lemma,
as 
\bea 
 \hbox { Number of ribbon graphs  }& =&  { 1   \over   4^v v!  } 
 \sum_{ \gamma \in  S_v [\mZ_4]  } \sum_{ \sigma \in [ 2^{2v} ] } \delta ( \sigma \gamma \sigma^{-1} \gamma^{-1} ) \cr 
& = &  { 1 \over |H_1|}  \sum_{ \gamma \in H_1 } \sum_{ \sigma \in S_{4v}/H_2 } 
        \delta ( \sigma \gamma \sigma^{-1} \gamma^{-1} ) 
\eea

Using the results of section \ref{vertexonwick}, we can also write 
\bea\label{vertexwickribbon} 
  \hbox{ Number of ribbon graphs } &= & { 1 \over 2^{2v}  (2v) ! } 
\sum_{ \gamma \in   S_{2v}[S_2]  } \sum_{ \sigma \in  [4^v]} 
\delta ( \sigma \gamma \sigma^{-1} \gamma^{-1} )  \cr 
& =&   { 1 \over |H_2| }  \sum_{ \gamma \in H_2 } \sum_{ \sigma \in  S_{4v}/H_1 }
 \delta ( \sigma \gamma \sigma^{-1} \gamma^{-1} ) 
\eea
There is also a way of writing this that is manifestly symmetric between exchange of $H_1, H_2$
\bea\label{multipls-count1} 
 \hbox{ Number of ribbon graphs } &= &  
{ 1 \over |H_1| |H_2| n! } \sum_{ R \vdash n } \sum_{\gamma_1 \in H_1     } 
\sum_{\gamma_2 \in H_2  } \chi_R ( \gamma_1 ) \chi_R ( \gamma_2) \cr    
 &= & \sum_{ R \vdash n } \cM^{ R }_{{\bf 1}_{H_1}} \cM^{ R }_{{\bf 1}_{H_2}}
\eea
Recall the notation  $\cM^{ R }_{{\bf 1}_{H_1}}$ for the multiplicity of the one-dimensional representation 
of $H_1$ when the irrep $R$ of $S_{4v}$ is decomposed into representations of the subgroup $H_1$.

\subsection{ How many ribbon graphs for the same Feyman graph ? } 

Ribbon graphs come equipped with a natural notion of a genus. For this reason
not all of the contractions leading to a given Feynman graph will produce the same
ribbon graph. As an example, consider vacuum graphs in $\phi^4$ theory with $v=1$
vertex. All three contractions give the same Feynman graph, but two
 of them give a 
genus zero ribbon graph and one gives a genus one ribbon graph : see Figure
\ref{fig:graphvertex}. 

The line of thinking developed above allows an elegant answer to the
 question of how 
many ribbon graphs are there for a given Feynman graph. 
In this case $ \Sigma_0$ can be chosen as a permutation made of 4-cycles 
\bea 
\Sigma_0 = (1234) (5678) ... (4v-3,4v-2,4v-1,4v)
\eea
The symmetry group of this permutation $S_{v} [ \mZ_4] $. 
This is the set of permutations in $S_{4v}$ which commutes with 
$\Sigma_0$. The Wick contractions are described by $\Sigma_1$ 
which are pairings, i.e permutations in the conjugacy class 
 $[2^{2v}]$ of $S_{4v}$. Ribbon graphs are orbits of  $S_{v} [ \mZ_4] $
action (by conjugation) on  the permutations in $[2^{2v}]$. 
We can first  decompose
 the set of all possible
 the pairings in  $[2^{2v}]$ into orbits of 
the group $S_v[S_{4}]$. Each such orbit is a Feynman graph. Then 
decompose each orbit 
into orbits of the subgroup $S_v[\mZ_4]$ of $S_v[S_{4}]$.
 The elements within one orbit of 
$S_{v}[S_4]$ will fall generically
 into multiple orbits of $S_{v}[\mZ_4]$, corresponding to multiple ribbon 
graphs. 
We can view the {\it genus} of the worldsheet described by the
ribbon graph as an invariant of the orbits of $ S_v [\mZ_4]$ acting 
on 2-cycles. 
If two 2-cycles are in the same orbit, 
they must be associated with the same genus. 
Consider the permutation obtained by multiplying $\Sigma_0$ 
 with any Wick contraction which belongs to a fixed orbit of 
 $ S_v [\mZ_4]$. Denote this permutation by $\sigma_3$ and the
 number of cycles in this permutation by $c$.
The genus $g$ of 
each $S_{v}[\mZ_4]$ 
orbit is related to the number of cycles in the permutation $c$ and 
number of vertices $v$ by
\bea
  2-2g=c-v
\eea 
This is obtained by applying the Riemann Hurwitz formula to 
maps the case of maps with three branch points determined by $\Sigma_0 , \Sigma_1, \sigma_3 $ \cite{dMRam}. 

It is straightforward to perform, in software such as GAP, 
 the decomposition 
into orbits of $S_v[S_4]$, and then refine the decomposition 
according to  $S_v[\mZ_4]$.

\begin{figure}[ht]
\begin{center}
\resizebox{!}{3cm}{\includegraphics{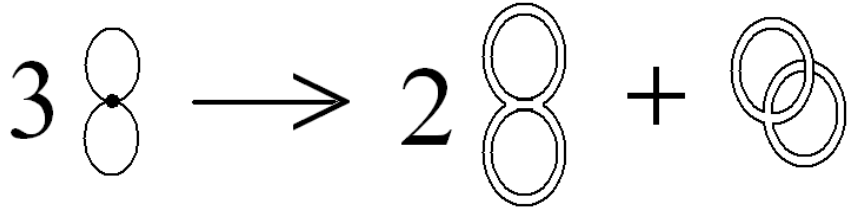}}
\caption{At $v=1$ the three possible Wick contractions give the same Feynman graph. Two of them give a 
genus zero ribbon graph and one gives a genus one ribbon graph.}
 \label{fig:graphvertex}
\end{center}
\end{figure}

A number of interesting directions can be contemplated.
Different ribbon graphs with the same underlying Feynman graph 
will have the same space-time integrals but possibly different 
genus. Our methods provide information about how the space-time 
dependence allows a certain range of genera. When we are dealing 
with vacuum graphs, each Feynman graph contributes a number. 
However, our techniques can be applied to graphs that have
external legs in which case we have explicit space-time dependences. 
Presumably a Feynman graph with a certain ``complexity'' - reflected 
in its space-time dependence, will allow a certain range of genera. 
The more complex it becomes, the more genera it will allow.
This could be studied quantitatively with the current set-up.

\subsection{ Using cycle indices for ribbon graphs } 

In this section we would like to count the number of ribbon graphs 
using cycle indices. 
For 4-valent ribbon graphs, the total number of graphs for $v$ vertices, 
is obtained by using the formula 
\bea\label{NformulaforRibbon}  
N ( ~ Z ( S_v [ \mZ_4 ] )  * Z ( S_{2v} [ S_2 ] )  ~ ) 
\eea
Thus, we need the cycle index of $S_v[\mZ_4]$. The cycle index of
any cyclic group is
\bea
Z^{\mZ_n}(\vec{x})={1\over n}\sum_{d\vdash n}\varphi(d)x_d^{n/d}
\eea
where $\varphi(d)$ is the Euler totient function. For $\mZ_4$ we find
\bea
Z^{\mZ_4}(\vec{x})={1 \over 4 } ( x_1^4 + x_2^2 + 2  x_4 )
\eea
Using known results for the cycle index of wreath products we now find
\bea 
\cZ^{ S_{\infty} [ \mZ_4 ] } [ t , \vec x ] \equiv
\sum_{n} t^n Z^{S_n[\mZ_4]}(\vec x ) 
= e^{ \sum_{i=1}^{ \infty}  { t^i \over 4i}  ( x_i^4  + x_{2i}^2  + 2 x_{4i})}  
\eea
It is now straight forward to obtain explicit answers for the number of ribbon graphs.

The methods at hand also apply to ribbon graphs with arbitrary numbers and types of 
traces. For a ribbon graph with $v_1 $ single traces, $v_2$ double traces, $v_3$ triple traces etc. we would use 
the group  $ S_{v_1} \times  S_{v_2}[Z_2 ]  \times S_{v_3} [  Z_3 ]  \cdots $ to count graphs.

\section{ QED  or Yukawa theory }\label{sec:QEDYUK} 

We have focused on real fields corresponding to particles which are neutral. In this section we explain
how our methods can be extended to the case of complex fields corresponding to charged particles. 
We will refer to the charged particle as an electron
with a view to applications to QED. Since 
are just counting Feynman graphs,  we do not track minus signs
coming from the anti-commuting nature of fermion fields. 
As far as the number of diagrams or their symmetry
factors goes, there is no difference between QED or Yukawa theory. Of course, in QED certain diagrams
vanish automatically due to Furry's Theorem. In the next section we will consider the problem of counting
the QED diagrams that remain after Furry's Theorem is applied.

The fact that we are dealing with charged particles implies that each edge will now have a preferred direction,
determined by the flow of charge. The model we have in mind has a cubic vertex in which the charged particle 
interacts with a neutral particle. This structure of the vertex matches both the Yukawa interaction and the coupling
of charged fermions to a gauge field. We will refer to the neutral particle as the photon. 
Thus, each vertex has 3 edges: an incoming electron, an outgoing electron and a photon. 
Consider vacuum Feynman graphs made of $2v$ vertices. 
We again clean the graph, by introducing new vertices in the middle of each edge. Label the vertices so that 
$1,...,2v$  are attached to the incoming electrons, $2v+1, \cdots,4v$ go with the outgoing electrons and 
$4v +1 \cdots 6v$ go with the photons. 

\begin{figure}[ht]
\begin{center}
\resizebox{!}{4cm}{\includegraphics{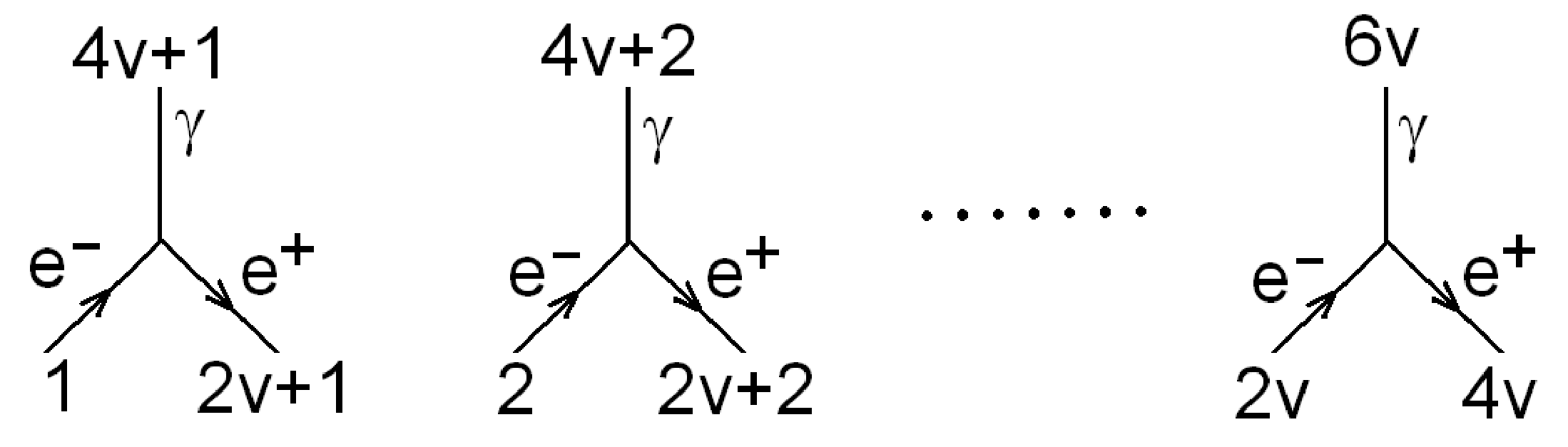}}
\caption{ QED vertices   }
\label{fig:QEDvertex}
\end{center}
\end{figure}

So we can describe the vertices by 
\bea 
\Sigma_0  = \prod_{i=1}^{2v}  < i , 2v+ i , 4v+i > 
\eea 
The Wick contractions are pairings of the form 
\bea 
\Sigma_1 =  \prod_{ i =1}^{2v} ( i , 2v + \tau_1 ( i ) ) \cdot 
\sigma_1 
\eea
The permutation $\sigma_1$ is in $[2^{v}]$ inside the $S_{2v}$ which acts on the last $2v$ among the 
$\{ 1 , \cdots , 6v \} $. The permutation $\tau_1$ is a general permutation of $2v$ objects. It tells us which 
of the first $2v$ (incoming electrons) go with which of the $2v+1 \cdots 4v$ (outgoing electrons). 
The symmetry $\gamma$ of $ \Sigma_0$ is just $S_{2v}$ of exchanging the $2v$ 
brackets. Because all legs are distinct there is no symmetry exchanging the legs at a vertex. 
Given the form of $ \Sigma_0$ we see that this is the diagonal $S_{2v}$ of the $S_{2v}\times S_{2v}\times S_{2v}$ 
acting on the electron, anti-electron and photon labels. We will call this ${\rm Diag}_3(S_{2v})$. 

The automorphisms of the Feynman graph are those elements $\gamma\in {\rm Aut}(\Sigma_0)={\rm Diag}_3(S_{2v})$
which also leave fixed the $\Sigma_1$ pairing determined by  $(\sigma_1,\tau_1)$. The order of this
group again determines the symmetry factor of the Feynman graph.

Distinct Feynman graphs are orbits of the permutations  
\bea
{\rm Diag}_3  ( S_{2v}) \rightarrow S_{2v} \times S_{2v} \times S_{2v}
\eea
acting on $ \Sigma_1 ( \sigma_1 , \tau_1) $, where the $\sigma_1,\tau_1$ are described above. 
We know that the stabilizer of $\sigma_1 \in [2^{v}]$ in $S_{2v }$ is
conjugate to $S_v[S_2]$.  This leads to  
\bea 
S_{2v} / S_v [S_2] = [2^v] 
\eea 
The permutation $ \tau_1$ mixes the first $2v$ incoming electrons with the next $2v$ outgoing electrons. 
The stabilizer of such a permutation is the diagonal $S_{2v}$ which simultaneously moves the 
first $2v$ with the next $2v$, so that $\tau_1$ is running over the coset 
\bea 
( S_{2v} \times S_{2v} ) / {\rm Diag}_2 ( S_{2v} )  
\eea
We use the subscript $2$ because it is the diagonal of the two $S_{2v}$.
This is enough to conclude that the Feynman graphs are in one-to-one correspondence with the 
points of the double coset 
\bea 
  {\rm Diag}_3  ( S_{2v})  \setminus 
( S_{2v} \times S_{2v} \times S_{2v} )  / ( {\rm Diag}_2 ( S_{2v} ) \times S_v[S_2] ) 
\eea
The ${\rm Diag}_2 $ is the diagonal of the first two $S_{2v}$ in $ S_{2v} \times S_{2v} \times S_{2v} $.

\subsection{ Double cosets of product symmetric groups } \label{dblcst}

To count the number of Feynman graphs, we need to count the number of points in the double coset 
\bea 
H_1 \setminus ( S_{2v} \times S_{2v} \times S_{2v} ) / H_2 
\label{qedcoset}
\eea
where $ H_1 = {\rm Diag}_3 ( S_{2v} ) $ and 
$H_2 = {\rm Diag}_2 ( S_{2v} ) \times S_v [ S_2 ] $. 
It is instructive to approach this problem 
by first considering cosets 
\bea
 H_1 \setminus G / H_2
\eea 
where $G$ is a general product of symmetric groups $ S_{ n_1} \times 
S_{n_2} \cdots \times S_{n_k} $, and $H_1 ,  H_2$ are general subgroups. 
Let us denote by 
\bea\label{gencase} 
\cN ( H_1 ,  H_2 ; G ) 
\eea
the number of points in this double coset.  The 
case $ \cN  ( H_1 , H_2 ; G = S_n ) =   N  ( H_1 * H_2 )$ 
is the one we already discussed in  sections 
\ref{sec:dubcoset} and \ref{Nformula}.  In the case at hand, we need
\bea 
\cN ( H_1 , H_2 ; S_{2v} \times S_{2v} \times S_{2v} ) 
\eea

As a consequence of the fact that the numerator group has changed,
when we compute the star product, it will turn out that 
we should multiply the coefficients 
of $ Z(H_1) , Z(H_2)$, and now weight them by the symmetry factors
appropriate for $S_{n_1 } \times S_{n_2} \times \cdots S_{n_k}$.
This is a generalization of Read's original formula \cite{Read} so it is worth discussing in some detail.

When the numerator group $G$ is a product of symmetric groups, 
a subgroup such as $H_1$ will have some number of elements in 
each conjugacy class of $G$, which is specified by three partitions. 
\bea
\{ p^{(1)} , p^{(2)} \cdots  p^{(k)} \} 
\eea
For the case at hand $k=3$. 
 The symmetry of such a conjugacy class is 
\bea 
{\rm Sym}(\{p^{(i)}\})=\prod_{i}{\rm Sym}(p^{(i)})=\prod_{i=1}^3\prod_{j=1}^{2v}j^{p_j^{(i)}} p_j^{(i)} !
\eea

We will restrict the discussion to $k=3$, but 
the generalization to any $k$ is immediate. 
For a subgroup $H$ of $G$ it is natural to define 
\bea 
Z^{ H \rightarrow G } = \sum_{ p^{(1)} , p^{(2)} , p^{(3)}  }  Z_{p^{(1)}, p^{(2)}, p^{(3)}    }^{ H\rightarrow G } \prod_ {i , j , k } x_i^{p^{( 1)}_i } 
   y_j^{ p^{(2)}_j   }  z_k^{   p^{(3)}_k   }   
\eea
where the coefficients  $Z_{p^{(1)}, p^{(2)}, p^{(3)}}^{ H\rightarrow G }$
keep track of the number of permutations in the subgroup with the cycle structure 
specified by the 3 partitions.
For two subgroups $ H_1 , H_2 $, we can use the two cycle indices (with respect to $G$) to define a 
generalization of the star product in (\ref{strry}) as follows 
\bea 
\cN ( H_1 ,  H_2 ; G ) = 
\sum_{ p } Z^{H_1 \rightarrow G }_{ p^{(1)} , p^{(2)} , p^{(3)} }  Z^{H_2  \rightarrow G }_{ p^{(1)} , p^{(2)} , p^{(3)} } 
 {\rm Sym} ( \{ p^{ (i)} \} )
\label{newread} 
\eea 

We can  understand this formula by adapting the reasoning in \cite{Read}. 
For any $ \rho \in G$,
the product $h_1\rho\ h_2$ gives $|H_1|\times |H_2|$ elements, that all
 belong to the same equivalence class in the  coset (\ref{gencase}),
as $h_1$ runs over $H_1$ and $h_2$ runs over $H_2$. Some elements will be repeated. When this happens
\bea
h_1\rho h_2 =\tilde{h}_1\rho \tilde{h}_2
\eea
which implies that
\bea
\tilde{h}_1^{-1}h_1 = \rho\tilde{h}_2 h_2^{-1}\rho^{-1}
\eea
Clearly $\tilde{h}_1^{-1}h_1\in H_1\cap \rho H_2\rho^{-1}$.
Given $\tilde{h}_1$, $\tilde{h}_2$ and an element of $H_1\cap \rho H_2\rho^{-1}$, $h_1$ and $h_2$ are uniquely determined. 
Consequently the number of times that $\rho$ is repeated, $\nu (\rho)$, is $|H_1\cap \rho H_2\rho^{-1}|$.  Every element in $G$ 
which appears in the equivalence class of $\rho$ comes with 
this same multiplicity in $ H_1 \rho H_2 $.
Hence,  the
number of distinct elements in the equivalence class of $\rho$ 
\bea
n(\rho)={|H_1||H_2|\over\nu (\rho )}
\eea
Consider   
\bea 
\sum_{ \rho \in G } \nu (\rho)
\eea 
For an equivalence class containing $\rho$, 
we get a contribution $ n(\rho)\nu (\rho)= |H_1||H_2|$. 
This is independent of the equivalence class. If there are $N_d$ equivalence 
classes, we get 
\bea
\sum_{ \rho \in G }  \nu (\rho) = N_d|H_1||H_2|
\eea
Rearranging we find
\bea
N_d={1\over |H_1||H_2|}\sum_{ \rho \in G }  \nu (\rho)
\eea
 To
compute the sum over $\rho$ of $\nu (\rho)$ we can choose $u_1\in H_1$ and $u_2\in H_2$ and count how
many $\rho$'s obey
\bea
u_1=\rho u_2\rho^{-1}
\eea
Since $u_1$ and $u_2$ are conjugate, they have the same cycle structure. Thus, to compute the sum over
$\rho$ of $\nu (\rho)$ we need to fix a cycle structure, multiply the number of elements in $H_1$ with this cycle
structure by the number of elements in $H_2$ with this cycle structure, then multiply by the number of elements of $G$ that
leave this cycle structure invariant, and then finally sum over cycle structure.
This is precisely what (\ref{newread}) is computing.

\subsection{ Analytic expressions for number of Feynman graphs in  QED } 

In this section we will apply the formulas obtained in the previous section to obtain
explicit results for the number of Feynman graphs. We need two cycle indices.
For $H_1={\rm Diag}_3 (S_{2v})$  we have 
\bea\label{zh1g} 
Z^{ H_1 \rightarrow G } = \sum_{ p \vdash 2v  } { 1 \over {\rm Sym}(p)} \prod_{ j=1}^{2v} ( x_j y_j z_j )^{p_j }  
\eea
For $H_2 = {\rm Diag}_2 ( S_{2v}) \times S_v [ S_2   ] $, we have 
\bea\label{zh2g} 
Z^{H _2 \rightarrow G } = 
\sum_{q\vdash 2v}\sum_{r \vdash 2v}{1\over{\rm Sym}(q)} Z_r^{ S_v [ S_2 ] } \prod_{ j=1}^{ 2v} ( x_j y_j )^{q_j }  
z_j^{ r_j }  
\eea

To calculate the $ \cN ( H_1 ,  H_2 ; G ) $, we need to multiply like terms in the (\ref{zh1g}) and (\ref{zh2g}), 
weighted by an appropriate symmetry factor. Picking up like terms forces $ q = p = r $ so that 
\bea\label{zh3g}  
\cN ( H_1 ,  H_2 ; G ) 
&& = \sum_{ p \vdash 2v } { 1 \over {\rm Sym}(p)}{Z^{S_v[S_2]}_p\over{\rm Sym}(p)}\cdot ({\rm Sym}(p)) ^3 \cr 
&&= \sum_{ p \vdash 2v } Z^{ S_v [ S_2  ] }_p {\rm Sym}(p) 
\eea

For the wreath product  
\bea 
Z^{ S_{\infty} [ S_2 ] } [ t  ; x_1  , x_2 , \cdots ]  & \equiv &  \sum_{ n=0}^{ \infty } Z^{ S_n [ S_2] } ( x_1 , \cdots , x_{ 2n} ) t^n  \cr 
& =& e^{ \sum_{ i =1}^{ \infty} {  t^i \over 2 i }  ( x_i^2  + x_{2i} ) }
\eea
To get the desired counting from this, we need to replace $ \prod_i x_i^{p_i} $ 
wth  $ i^{p_i} p_i!$. Equivalently, do a replacement $ x_i \rightarrow i y_i $ 
and expand in $y_i$ replacing $y_i^{p_i}$ with $p_i!$. In terms of 
\bea 
\tilde Z  [ t  ; y_i  ]
& = &  Z^{ S_{ \infty} } [ t  ; x_i = i y_i  ] \cr 
& = &  e^{ \sum_{ i =1}^{ \infty} {  t^i \over 2 i }  (i  y_i^2  + 2  y_{2i} ) }
\eea
the counting of vacuum graphs with $2v$ vertices can be written as 
\bea 
\cN  ( H_1 ( v) ,  H_2 (v)   ; G ( v ) ) = \oint {  t^{- 2v} dt
\over t }  \bigl ( \prod_i \int_0^{ \infty} dy_i e^{ - y_i    } \bigr )  \tilde Z [ t ; y_1 , y_2 \cdots ]  
\label{QEDCOUNT}
\eea
See Appendix \ref{data} for explicit numerical results obtained using these formulas.

\subsection{ QED and TFT for product symmetric groups } 

The number of Feynman graphs is counted by the formula
\bea 
{ 1 \over |H_1| | H_2|  |G |  } \sum_{ u_1 \in H_1 } \sum_{ u_2 \in H_2  } \sum_{ \gamma \in G } \delta ( u_1 \gamma u_2 \gamma^{-1} ) 
\eea
We have already established that
\bea 
G & = &  S_{2v} \times S_{2v} \times S_{2v} \cr
H_1 & = &  {\rm Diag}_3 ( S_{2v} ) \cr 
H_2 & = &  {\rm Diag}_2( S_{2v} ) \times S_{v} [ S_2 ] 
\eea

This is a cylinder partition function in $ S_{ \infty} \times S_{\infty} \times S_{\infty} $ gauge theory 
which is Schur-Weyl dual to the large $N $ limit of $ U ( N ) \times U ( N) \times U ( N) $ 
gauge theory. As before the correspondence holds in the zero area limit.

\section{ QED with the Furry's theorem constraint }
\label{sec:QEDfurry}  

Above we have counted the total number of Feynman diagrams in QED or Yukawa theory. In the case of
QED, Furry's Theorem proves a fermion loop punctuated with an odd number of photons vanishes \cite{peskin}. In this 
the number of  QED Feynman graphs, not vanishing by the Furry constraint, 
are counted. Enumeration of some Feynman graphs with their 
symmetry factors for this case is given in \cite{cvitan2}.

\begin{figure}[ht]
\begin{center}
 \resizebox{!}{6cm}{\includegraphics{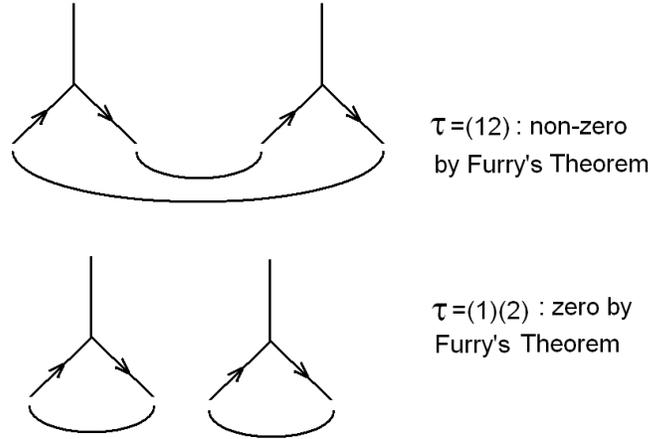}}
\caption{ Even cycle lengths and Furry's theorem }
 \label{fig:QEDfurry}
\end{center}
\end{figure}

\subsection{ Furry's theorem and a constraint to even cycles   } 
Following the discussion of the last section, each QED Feynman graph is specified by 
\bea 
\Sigma_0 & =  & \prod_{i=1}^{2v} < i ;  2v + i ;  4v + i > \cr 
 \Sigma_1 ( \sigma_1 , \tau_1 ) & =&  
\prod_{ i=1}^{2v} ( i , 2v + \tau_1 ( i) )  ~ \cdot ~   \sigma_1  
\label{qqedpair} 
\eea
Recall that $\tau_1$ specifies how incoming and outgoing electrons are connected.
Consequently, the number of cycles appearing in $\tau_1$ is equal to the number of 
fermion loops in the graph and the length of each cycles is equal to the number of
photons decorating the loop. It is clear that 
the constraint that each fermion loop has only an even number of photons 
is easily imposed in the $(\Sigma_0 ,\Sigma_1 )$ description by requiring 
that $\tau_1$ consists of permutations which only have cycles of even length, as shown in Fig 20. 

The orbits of the group $ G = S_{2v} \times S_{2v} \times S_{2v}$ acting on the pairs $(\Sigma_0 ,\Sigma_1)$
correspond to the inequivalent Feynman graphs. 
 The set of $ \gamma \in G $
which preserve the pair $( \Sigma_0 , \Sigma_1 )  $ generates 
the automorphism group of the graph and the order of this 
automorphism group is the symmetry factor. Preserving 
$ \Sigma_0 $ forces $ \gamma $ to be in the diagonal $S_{2v}$ 
of $ G$ : 
\bea\label{diagact}  
 \hbox{For }  1 ~ \le ~  i & \le &  2v\cr 
  i & \rightarrow & \gamma (i )  \cr 
 2v + i & \rightarrow &  2v + \gamma (i) \cr 
 4v + i & \rightarrow & 4v + \gamma (i ) 
\eea
We can write this as $\gamma \circ \gamma \circ \gamma $,
where the first $\gamma$  acts on the first $2v$ according to the second line 
of (\ref{diagact}), while  leaving 
 the subset $\{2v+1, \cdots, 6v\}$ fixed
; the second  $\gamma $  moves only the 
subset $ \{ 2v+1 , \cdots , 4v \} $ according to the 3rd line; 
and  the third $ \gamma   $  moves only the 
subset $ \{ 4v +1 , \cdots , 6 v \} $ according to the 4th  line. 

The symmetry factor of a Feynman graph specified by 
the standard $ \Sigma_0 $ and a general $ \Sigma_1 $ is 
\bea 
|{\rm Aut}(\big[\Sigma_1\big]_{H_1})|=\sum_{\gamma \in S_{ 2v }  } 
\delta (  \gamma \circ \gamma \circ \gamma  
 ~ \Sigma_1  ~  \gamma^{-1}  \circ \gamma^{-1} \circ \gamma^{-1} 
  \Sigma_1^{-1} ) 
\eea
$ \Sigma_1 $ is not a  permutation in $ G$, since the $\tau_1 $
mixes the first $2v$ with the second $2v$. 
The multiplication can be viewed in $S_{6v}$ or $ S_{4v} \times S_{2v}$

A simple application of the Burnside Lemma implies that the number of Feynman graphs
remaining after Furry's Theorem is applied is 
\bea 
{ 1 \over( 2v)!  }\sum_{\gamma \in S_{ 2v }  } \sum_{ \sigma_1 \in [2^{v}]  \in  S_{2v}   } ~~  
\sum_{ \tau_1 \in S_{2v} : [ \tau_1 ] \hbox{ even }  } 
 \delta ~ ( ~~  \gamma \circ \gamma \circ \gamma ~~   \Sigma_1 ( \sigma_1 , \tau_1 )   \gamma^{-1} \circ \gamma^{-1} \circ \gamma^{-1}  ~~  \Sigma_1^{-1} ( \sigma_1 , \tau_1 )  ~~  ) 
\cr 
\eea

Another approach to the same counting problem would be to implement the Furry constraint
in the double coset language. Recall the double coset relevant for QED is 
\bea\label{doubcosetforqed}  
  {\rm Diag}_3( S_{2v} )  \setminus ( S_{2v} \times S_{2v} \times S_{2v} ) / {\rm Diag}_2 ( S_{2v} ) \times S_v [ S_2 ]  
\eea
Using (\ref{deltaRead}) with the $G, H_1, H_2$ identified according to 
(\ref{doubcosetforqed}) the size $F$ of this double coset is counted by 
\bea 
&&F =  { 1 \over (2v)! (2v)!  2^v v! } \sum_{ \sigma \in S_{2v} }  \sum_{ \tau \in S_{2v} } \sum_{ \rho \in S_v [ S_2 ] } 
\sum_{b_1  , b_2 , b_3 \in S_{2v }  } \cr  
&& \delta_{ S_{2v} \times S_{2v} \times S_{2v} }  ( \sigma \circ  \sigma \circ  \sigma ~~ b_1 \circ  b_2 \circ b_3 ~~ 
             \tau \circ \tau \circ \rho  ~~    b_1^{ -1}  \circ  b_2^{-1}  \circ b_3^{-1} ) 
\eea
$b_1$ acts on the incoming electrons, $b_2$ on the outgoing electrons and $b_3$ on the photons.
$\sigma$ permutes vertices in the graph, $\tau$ permutes electron lines and $\rho$ permutes photon lines. 
This can be simplified to
\bea 
&& F =  { 1 \over (2v)! (2v)!  2^v v! } \sum_{\sigma\in S_{2v}}  \sum_{\tau\in S_{2v}} \sum_{\rho\in S_v[S_2]} 
\sum_{b_1  , b_2 , b_3 \in S_{2v }  } \cr  
&& \delta_{ S_{2v} \times S_{2v} } (  \sigma \circ  \sigma ~~ 
  b_1 \circ  b_2 ~~\tau \circ \tau ~~  b_1^{ -1}  \circ  b_2^{-1} ) \cr 
&& \delta_{ S_{2v} } (\sigma  b_3  \rho b_3^{-1}   ) 
\eea
Solving the second delta function for $\sigma$, we can plug back into the
first delta function and do a redefinition of the sums $ \sum_{b_1, b_2 } $
to absorb the $b_3$, so as to get 
\bea\label{absb3}  
 F & = &  { 1 \over  (2v)!  2^v v! }  \sum_{ \tau \in S_{2v} } \sum_{ \rho \in S_v [ S_2 ] } 
\sum_{b_1  , b_2  \in S_{2v }  } \cr 
&& \qquad \qquad \qquad \delta_{ S_{2v} \times S_{2v} } 
(\rho\circ\rho ~~ b_1\circ b_2 ~~\tau \circ \tau ~~  b_1^{ -1}  \circ  b_2^{-1} )\cr 
& = &  { 1 \over  (2v)!  2^v v! }   \sum_{ \rho\in S_{2v} } \sum_{ \tau_1 \in 
S_v [ S_2 ] } 
\sum_{b_1  , b_2  \in S_{2v }  }
 \delta_{ S_{2v}} (\rho b_1 \tau  b_1^{ -1}  )
  \delta_{ S_{2v}} (\rho b_2 \tau  b_2^{ -1}   ) \cr 
&=&  { 1 \over  (2v)!  2^v v! }  \sum_{\rho\in S_v [ S_2 ] } 
\sum_{b_1  , b_2  \in S_{2v }  }  
 \delta_{ S_{2v}} ( b_2^{-1}\rho^{-1} b_2 b_1^{-1}\rho  b_1) \cr 
& =&  { 1 \over    2^v v! } \sum_{ \rho \in S_v [ S_2 ]  }
 \sum_{\tau_1 \in S_{2v} }
\delta_{ S_{2v} } (\tau_1 \rho \tau_1^{-1} \rho^{-1}   ) 
\eea 
In the last line we have recognized that, given the interpretation of $b_1,b_2$ it is clear that
$\tau_1=b_1^{-1}b_2$ is the permutation $\tau_1$ in (\ref{qqedpair}). 
The Furry constraint is now easily implemented. 
Thus, the number of Feynman graphs remaining after the Furry constraint is 
implemented is given by 
\bea\label{numfeynfurr} 
&& \hbox{ Number of Feynman graphs for QED with Furry constraint  } \cr 
&& =  { 1 \over    2^v v! } \sum_{\tau_1 \in S_{2v} : even} ~~ 
 \sum_{ \rho  \in S_v [ S_2 ]  }
\delta_{ S_{2v} } (  \tau_1 \rho \tau_1^{-1}\rho^{-1} )  
\eea
The permutation $\tau_1$ is constrained to have even cycles only and
we are summing over elements of $S_{v}[S_2]$. 
For each element of $S_v[S_2]$, the weight is the number of permutations with even cycles only, 
which commute with the given permutation in $S_v[S_2]$. 

From the first line of (\ref{absb3}) we see that an equivalent double coset description of the 
counting is 
\bea 
 S_{2v}   \setminus ( S_{2v} \times S_{2v})  / {\rm Diag}_2 ( S_v[S_2] )  
\eea
This gives a slightly simpler connection to 
observables for 2d Yang-Mills  with cylinder target space. 
Namely we have a connection to $U(\infty )\times U(\infty )$ 
rather than $U(\infty )\times U(\infty )\times U(\infty )$
as described earlier.  

\subsection{ From QED to ribbon graphs }\label{QEDtoribbon} 

Comparing the expressions (\ref{absb3}) and (\ref{numfeynfurr}) 
with (\ref{multipls-count1}) for ribbon graph counting, it becomes 
clear that the counting of QED Feynman graphs can be matched 
with counting ribbon graphs. The restriction to vertices 
of valency $4$ in  (\ref{multipls-count1}) is being relaxed to 
allow arbitrary valencies in  (\ref{absb3}) and arbitrary 
even valencies in  (\ref{numfeynfurr}). 
We will now  explain  the bijection between ribbon
graphs and QED/Yukawa graphs which explains the equality of 
counting formulae.

\begin{figure}[ht]
\begin{center}
 \resizebox{!}{6cm}{\includegraphics{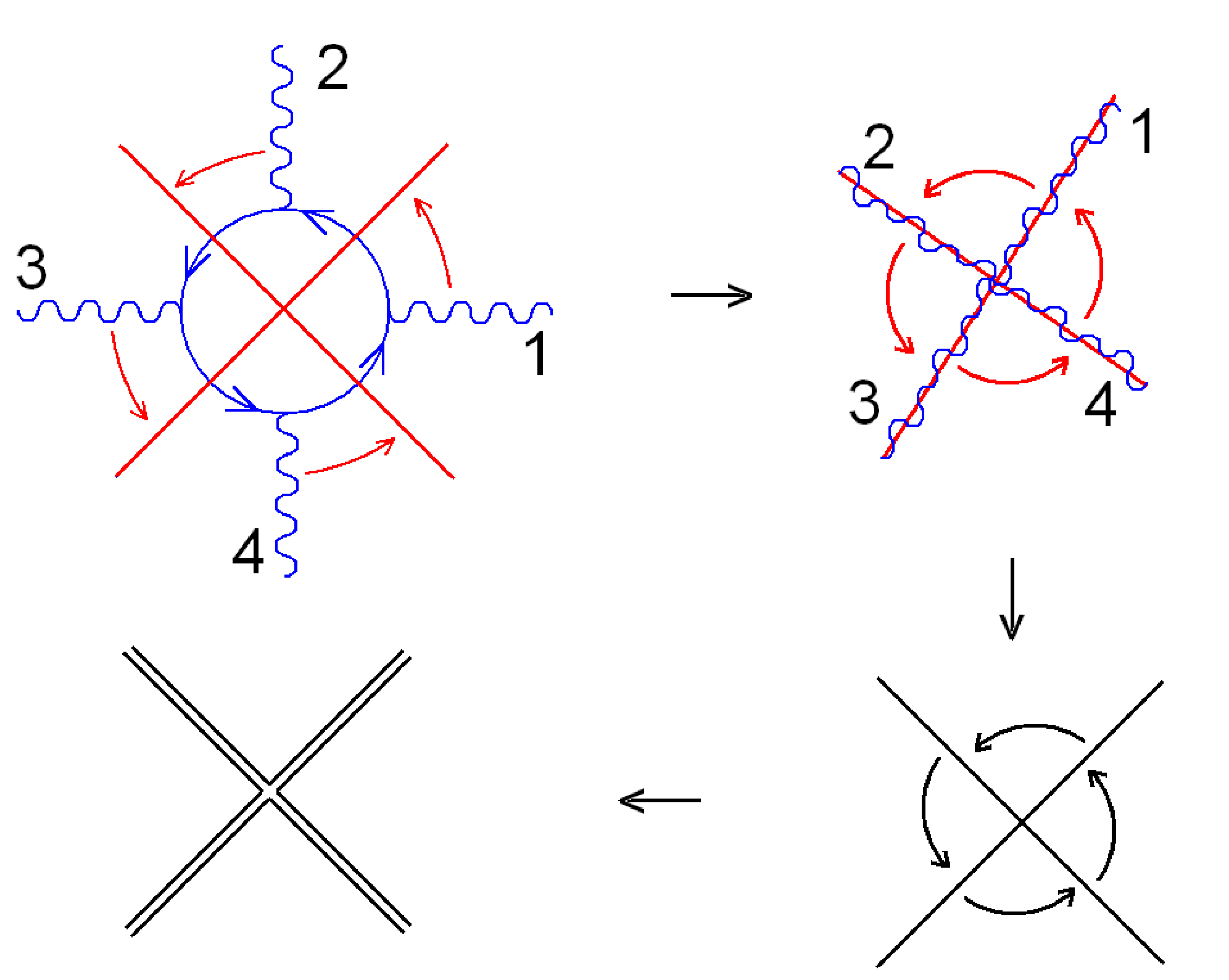}}
\caption{QED graphs to ribbon graphs  }
 \label{fig:Ribbonvertex}
\end{center}
\end{figure}

The general QED/Yukawa vacuum graph has loops with vertices 
where the photon joins the 
loop. These loops determine $2v$ cycles which form a permutation in $S_{2v}$. 
To build a bijection to ribbon graphs think of the photon 
labels in Figure \ref{fig:QEDvertex} as attached to the 
vertices. The photon contractions determine an element of $[2^{2v}]$. 
At the centre of each electron loop draw a point and radiate edges (spokes)  
to intersect the edges of the loop, one spoke between each pair of 
electron-photon vertices. Use the arrow on the electron propagator to 
move each end of the photon propagator along the electron loop 
towards the intersection of the spoke. Erase the electron loop, leaving 
the spokes and the photon propagators connecting them. This gives a 
graph with vertices of arbitrary valency, but with the vertices 
being equipped with a cyclic order, which is nothing but a ribbon graph,
completing the construction of the bijection. This process is illustrated
in Figure \ref{fig:Ribbonvertex}. The bijection demonstrates that the number of QED vacuum 
graphs with $2v$ vertices is the same as the number of ribbon graphs with 
$v$ edges (the photon propagators become edges). 

This means that the total valency of the vertices 
in the ribbon graph is $2v$. Each vertex can have any 
integer valency compatible with this constraint. Thus, following 
 the discussion (\ref{vertexwickribbon}) and (\ref{NformulaforRibbon}),   
 the number of QED/Yukawa vacuum graphs with $2v$ vertices is
\bea 
\hbox{ Number of Feynman graphs for QED  } = 
\sum_{ p \vdash 2v }  N ( Z(H_1)  * Z(H_p) ) 
\eea
where $p$ is a partition of $2v$, the group $H_1$ is $S_v[S_2]$ and the group $H_p$ is 
\bea 
S_{p_1} [ \mZ_1  ] \times S_{p_2} [ \mZ_2 ] \times \cdots S_{p_{2v} } [ \mZ_{2v}  ]  
\eea
The function $N$ of two cycle index polynomials is as defined 
in (\ref{strry}).
For QED with the Furry constraint implemented, the valencies are constrained to be even
so that
\bea
&& \hbox{Number of Feynman graphs for QED with Furry constraint} \cr 
&& = 
\sum_{ \substack{ p \vdash 2v  \\
         p  ~~ even } }  N ( Z(H_1) * Z(H_p) ) 
\label{FurryQED}
\eea

Given the connection between ribbon graphs and Belyi pairs \cite{dMRam}
we see that QED counts the number of clean Belyi pairs 
(bi-partite embedded graphs with bivalent white vertices) 
with degree $v$ and black vertices of any valency. 
QED with the Furry constraint counts the number of such Belyi pairs 
with black vertices of even order only.

\section{ Discussion  }\label{sec:discussion}

\subsection{ Permutations and Strings : a new perspective on  
   gauge-string dualities.  } 

We have seen that the counting of Feynman diagrams 
can be interpreted in terms of string amplitudes 
with a cylinder target space. The nature of the 
interaction vertices determines a permutation symmetry group 
$H_1$ and the Wick contractions determine a permutation symmetry group 
$H_2$. The counting could also be written in terms of 
commuting pairs of permutations, where one permutation 
lives in one of these subgroups, and another one lives 
in a coset involving the other group. This latter realization 
can be interpreted in terms of covering maps of a  torus target space, 
such as those that arise in the string theory of 
2d Yang Mills for torus target space \cite{gt1,gt2}, except 
that there is a constraint on the two permutations 
associated with the cover, by following the inverse image of 
paths along the $a$ and $b$ cycle. 

In 2d Yang Mills, one encounters a sum over commuting permutations $s_1, s_2$
which is equivalent to counting the set of all covers of the torus
by a torus. This sum is invariant under the $SL(2,Z)$ of the  
space-time torus, which acts on these permutations by 
\bea 
 S : \begin{pmatrix} s_1 \\ s_2 \end{pmatrix} && \rightarrow  
\begin{pmatrix} s_2  \\  s_1^{-1}  \end{pmatrix} \cr 
 T :  \begin{pmatrix} s_1 \\ s_2 \end{pmatrix} && \rightarrow  
     \begin{pmatrix} s_1s_2 \\ s_2 \end{pmatrix} 
\eea
and obeys the $SL(2,Z)$ relations : 
\bea 
 S^2  \begin{pmatrix} s_1 \\ s_2 \end{pmatrix} && \rightarrow   \begin{pmatrix} s_1^{-1}  \\ s_2^{-1}  \end{pmatrix} \cr 
 (ST)^3  \begin{pmatrix} s_1  \\ s_2  \end{pmatrix}
 && \rightarrow  
\begin{pmatrix} s_1  \\ s_2 \end{pmatrix}
\eea
In the case at hand, the $SL(2,Z)$ of space-time is 
broken in that $s_1 $ is summed over $H_1$ while $s_2$ is 
summed over $ S_{4v}/ H_2 = [ 2^{2v} ] $.

While making two choices of observable at each end 
of a cylinder seems like a natural construction for string
theory with a cylinder, the constraints on the covering maps 
of the torus look rather intricate, and are not of any sort 
that has been discussed in the literature on topological strings. 
Nevertheless, there might well be a string construction 
which sums over maps with constraints of this sort. 

In any case,  the Feynman diagram combinatorics 
suggests two emergent dimensions of cylinder or torus, 
with one formulation possibly more appealing than the other. 
In the construction of correlators of general trace operators
in the Hermitian Matrix model, there is another 2-dimensional 
target space emerging, which is the sphere with three 
punctures \cite{dMRam}. 

In the application of Feynman graphs to QFT, the 
Feynman graph counting is only part of the  answer. 
When there are external edges, each edge is associated with a 
spacetime position (or momentum), each of which takes values in $\mR^4$.
The computation of Greens functions $ \langle \phi(x_1) \phi(x_2) \cdots \phi (x_E ) \rangle $ is a sum over Feynman diagrams, which live in 
\bea\label{dc} 
 DC ( v , E )  =  ( S_{v }[S_4] \times S_1^E )     \setminus   S_{4v+E } / S_{2v+E/2} [ S_2] 
\eea
This has an action of $S_E$. Projecting down the $S_E$ orbits 
gives another double coset. There is a fibration 
\bea 
&&  ( S_{v }[S_4] \times S_1^E )     \setminus   S_{4v+E } / S_{2v+E/2} [ S_2] 
   \leftarrow S_E \cr 
&& \qquad \qquad \downarrow  \cr 
&&  ( S_{v }[S_4] \times S_E )     \setminus   S_{4v+E } / S_{2v+E/2} [ S_2]
\eea
We may write 
\bea 
G( x_1 , x_2 , \cdots , x_E ) = \sum_{ \Sigma  } 
G ( x_1 , x_2 , \cdots x_E ; \Sigma )   
\eea
where $ \Sigma $ lives in the double coset (\ref{dc}). 
The Bose symmetry of $G( x_1 , x_2 , \cdots , x_E ) $ 
arises as an $S_E$ invariance of the summands
\bea 
G ( x_{\sigma(1)}  , x_{ \sigma(2)}  , \cdots x_{ \sigma ( E) }  ; 
\sigma ( \Sigma )  )  
\eea
This shows that in some sense 
the space-time coordinates $(\mR^4)^E$ and the double coset $DC( v , E ) $  
associated with covers of a cylinder may usefully be viewed
as being on an equal footing when we work out the physics. 
It is tempting to infer that this is indicative of an underlying 
$ \mR^4 \times \hbox{ Cylinder}  $ or $ \mR^4 \times\hbox{Torus} $
in four dimensional QFTs. We have shown that  the extra two dimensions emerge
 from the Feynman graph combinatorics, but it remains to be seen 
if there is a dynamical six dimensional picture along these lines, 
and how it might relate to other appearances of six dimensions 
in 4D QFTs such as \cite{wittenMHV}.

A very interesting problem is to interpret the full correlators or 
S-matrices, in terms of the cylinder (or torus) emerging from 
combinatorics, with the $\mR^4$ which is manifestly there to begin 
with. This would be an example of a holographic dual of 
a quantum field theory, analogous to AdS/CFT, where 
from the point of view of the dual gauge theory, 
an emergent $S^1$ along with an emergent radial direction, resulting 
in $M_5 \times S^1$, with the original $\mR^4$ of the gauge theory 
realized as the boundary of the $M_5$. 

If this is indeed possible with $\phi^4$ theory, 
it would be an example of gauge-string duality 
with an interesting difference from all the examples 
known so far. It would not rely on the large $N$ expansion, 
and ribbon graphs thickening into string worldsheets. 
Rather it would be a realization through the fundamental 
link between permutations and strings, of the physical 
expectation of holography   \cite{tHooft,susskind}.
This makes it very important to know whether 
this works or not. For some general recent discussions of 
the scope of holography see \cite{howkPap}.  
Another direction for relating Feynman graphs to string 
amplitudes, by directly identifying the integrand on 
the moduli space of complex structures of string worlsheets, 
has been advocated in \cite{gopakprog}.  The ideas emerging 
from the permutations-strings connection may well have interesting overlaps 
with this programme, which we will leave to future investigation.

If there are indeed new full-fledged gauge-string dualities 
for theories such as $\phi^4$ in four dimensions, or $\phi^3$ 
in six dimensions or indeed QED or QED-like theories, 
there are some obvious questions to be addressed. 
What is the map between physical parameters? 
In the $\phi^4$ theory, Feynman graphs with $v$ vertices 
are weighted with $g^v$. We have associated them with 
cylinder worldsheets covering a target space cylinder, 
with maps of degree $4v$. If the cylinder has area $A$,
the Nambu-Goto action will weight such strings with $e^{-nA}$. 
This suggests that the QFT coupling constant $g$ should 
be identified with $e^{-A}$. 

Another obvious question : What is the string coupling 
of this dual string theory? Since all the details of the 
field theory interaction translate into the group $H_1$ 
at a boundary of  the cylinder amplitude, interaction in the field 
theory do not translate directly into interactions of the string. 
A related question is whether there is some modification 
of the Feynman graph counting problem on the field theory side 
which leads to higher genus string worldsheets covering the cylinder.
 This requires 
further exploration of the sort of string amplitudes that come from 
various QFT Feynman diagrams. The added motivation for  these investigations 
is that the stringy picture leads directly to powerful counting 
results, such as those in Appendix D.

\subsection{  Topological Strings and worldsheet methods } 

 We have argued that there is a link between Feynman graph counting 
 and string amplitudes by recognizing the 
 combinatorics of worldsheet maps to a cylinder in the 
 Feynman graph combinatorics. The string theory in question 
involves the one that appears in the large N expansion 
of 2dYM with $U(N)$ gauge group.
This string theory also has a spacetime description 
in terms of topological lattice theory with $S_n$ (for all $n$, hence 
probably better described as $ S_{\infty}$) as gauge group. 
The observables needed to describe 
Feynman graph combinatorics are slightly more general 
than the ones that appear in $U(N)$ 2dYM.  They involve boundary observables 
which are not invariant under the entire $S_n$ in the lattice TFT 
description, but only certain specified subgroups $H_1,H_2$. 

Several worldsheet approaches to the string theory of 2dYM 
have been proposed  \cite{CMR,horava,vafa,AOSV,szabo2dYM}. 
 It will be very interesting to 
 develop, in   these approaches, 
 the boundary observables which count the Feynman graphs
and their symmetries. The geometrical description 
in terms of gluing world-sheets in section \ref{sec:symmfacstrings}
should be useful. An interesting goal would be to 
find new  worldsheet methods for calculating Feynman graph 
combinatorics, which may be efficient for obtaining asymptotic results.

\subsection{ Collect results on orbits } 

We have developed a method to think about the combinatorics 
of Feynman graphs,  involving pairs $(\Sigma_0 , \Sigma_1 ) $ 
of combinatoric data. This was discussed from the point of 
view of double cosets, which provide a link to generating functions. 
From the point of view of constructing the Feynman graphs,
 perhaps the most immediate lesson  extracted from the 
study of the pairs  $(\Sigma_0 , \Sigma_1 ) $ is that 

{\vskip 0.5cm}

\fbox{  
    \begin{minipage}[c]{14cm}
    \flushleft{A Feynman graph is an orbit of the vertex symmetry group
               acting on the Wick contractions.}
    \end{minipage}
}

{\vskip 0.5cm}

It is a simple exercise to obtain both the vertex symmetry group and the set of Wick contractions
given a quantum field theory. We have listed this data in the table below for the theories that
we have considered here. The data is relevant to vacuum graphs - there are 
simple generalizations including external edges. 

\begin{center}
\begin{tabular} { |l | l  | l | l|  }  
\hline
Group & acts on & Feynman Graph problem  \\ \hline
$S_v[S_4]$  & permutations in  $[2^{2v}]$ of  $S_{2v} $  & $\phi^4$ theory  
\\ \hline 
$S_v[S_3]$ &  permutations in  $[2^{3v\over 2 }]$ of  $S_{3v/2} $ &  $\phi^3$ theory  \\ \hline 
$ S_{v_4} [S_4]  \times S_{v_3} [S_3]$ &
  permutations in  $[2^{( 3v_3 + 4v_4 ) \over 2 }]$ of  $S_{3v_3  + 4v_4}  $
& $ \phi^3 + \phi^4$ theory \\ \hline 
 $S_{v} [ S_2 ]$ &  All permutations in $S_{2v}$ & Yukawa/QED  \\  \hline 
 $S_{v} [ S_2 ]  $ & All even-cycle permutations in $S_{2v}$ & Furry QED  \\ \hline 
 $S_{v} [ \mZ_4 ]$  &  permutations in  $[2^{2v}]$ of  $S_{2v} $ & Large $N$ expansion of Matrix $\phi^4$  \\  
\hline
\end{tabular} 

\end{center} 

The double coset picture  allows us to 
recognize  that one can also consider the symmetry of 
the Wick contractions (e.g $S_{2v } [S_2]$ in case of $\phi^4$) 
acting on the cosets associated with the vertices 
($S_{4v}/S_v[S_4] $ in this case).

\subsection{ Galois Group and Feynman Graphs } 

We showed that the counting of vacuum graphs in 
QED/Yukawa and QED with Furry constraint implemented 
can be mapped to counting of 
ribbon graphs. In the case of QED with Furry constraint, there is an even-only 
restriction on the vertices. We know, from the theory of Dessins d'Enfants 
\cite{grot},
 that there is an action of the absolute Galois group 
$Gal ( \bar { \mathbb{Q}}   / \mathbb{Q} ) $ on 
ribbon graphs. (For recent applications of Dessins d'Enfants
to Hermitian Matrix Model correlators, supersymmetric gauge theories
 and extensive references to the Mathematical literature see 
\cite{cach,dMRam,jrr,hhjprr}.)  
This means that there is an action of the Galois group on 
the QED vacuum diagrams. The list of orders of the vertices 
is a Galois invariant \cite{jones}, so the Furry theorem restriction continues 
to allow the Galois action to close. 

A related fact is that sums  over permutations restricted by conjugacy classes
are known to be sums over Galois group orbits. 
\bea 
\sum_{ \sigma_1 \in T_1 } \sum_{ \sigma_2 \in T_2 }  
\sum_{ \sigma_3 \in T_3 } \delta ( \sigma_1  \sigma_2 \sigma_3  ) 
\eea

Given the link we have observed between counting ribbon graphs 
and counting QED graphs, we have Galois actions 
on the QED/Yukawa and Furry QED vacuum graphs. 
We do not know if such  is the case for the vacuum graphs 
of scalar field theories. The answer depends on whether 
sums of the following form 
\bea 
\sum_{ \sigma_1 \in H_1 } \sum_{ \sigma_2 \in H_2 }  
\sum_{ \sigma_3 \in S_n  } \delta ( \sigma_1  \sigma_3^{-1} \sigma_2 \sigma_3 )   
\eea
for $H_1 = S_{v}[S_4 ] , H_2 = S_v [ S_2] $ are related to Galois theory. 
We know if $H_1 =  S_{v}[\mZ_4 ]$, because the connection between QED 
and ribbon graphs we described, these sums count Dessins. For
$H_1 = S_v [ S_4] $ we do not know an argument to relate to Galois 
actions.

\section{  Summary and Outlook   }\label{sec:summout}  

We have shown that the counting of Feynman graphs  and their
 symmetry factors
in scalar field theory, e.g $ \phi^4$ or $\phi^3$ 
theories, as well as QED, can be mapped to string amplitudes. 
The string amplitudes are of the type that appear in the 
string dual of large N Yang Mills theory, for which there are 
 several proposed worldsheet constructions. The large $N$ 2dYM 
observables are known to be related to $S_n$ TFT. 
The counting problems related to Feynman graphs have been expressed 
as observables in this $S_n$ TFT.  We used this $S_n$ TFT 
data to construct covers of the cylinder. The covers are inrepreted 
as string world-sheets, which are also of cylinder topology. 
 The form of the interactions in the QFT 
determines the observables at the two ends of the 
spacetime cylinder, which constrain the windings of the 
string worldsheets to belong to certain subgroups of
the permutation group.

The formulation in terms of string amplitudes 
is directly related to some classic formulae 
on counting of graphs in papers by Read\cite{Read}. 
We have found it useful to think about these 
formulae in terms of an operation of introducing 
an additional vertex in the middle of each edge of 
the Feynman graph. This separates each edge into a
pair of half-edges. We label the half-edges with numbers 
$ 1 , 2 , \cdots , n   $, and associate a quantity $ \Sigma_0$ 
describing the vertices of the Feynman graph. The newly-added 
vertices are described by a quantity $\Sigma_1$.
This operation is called ``cleaning'' in the context of 
ribbon graphs and related Belyi maps. In that case 
two permutations $ \sigma_0, \sigma_1$ play the role 
of $\Sigma_0, \Sigma_1$. In the case at hand, $ \Sigma_0$ 
is generically not a permutation, but can be viewed as 
labeling a coset of permutation groups.

There are groups $H_1$, $H_2$ which are symmetry groups
of $\Sigma_0$ and $ \Sigma_1$ respectively. The counting 
of Feynman graphs is equivalent to the counting of elements
in the double coset of the form 
\bea 
   H_1 \setminus \hbox{ ( Permutation group )} / H_2 
\eea
For the case of scalar field theories the permutation group 
is something of the form $S_n$. In QED, it is a product group.

The double coset connects directly to a string amplitude 
with cylinder target space, where $H_1,H_2$ are associated 
with the boundaries. The similarities in the current
approach between Feynman graphs and ribbon graphs, allows us to 
formulate in group theoretic terms, and derive nice formulae 
for questions such as the total number of types 
of ribbon graphs (summed over genus). Further it allows us 
to express as a group theory problem the counting of the number of 
ribbon graphs which correspond to the same Feynman graph. 
We may interpret this by saying that large $N$ ribbon graphs (say of 
matrix $\phi^4$ theory) arise from ordinary  Feynman graphs 
(of $\phi^4$ theory) by a symmetry breaking of $S_{v} [ S_4] $ 
to  $S_{v} [ \mZ_4] $.

The formulae we obtain for QED Feynman graph counting 
turn out, upon simplification,  to be related to the counting 
of ribbon graphs. We give an explanation of this relation 
by mapping QED graphs to ribbon graphs. The key point 
is the cyclic nature of ribbon graph vertices which 
map to orientations of fermion loops, in the correspondence 
we describe between ribbon graphs and QED graphs. 

We outline some avenues for  extensions of this work. 
It will be desirable to understand in terms of double cosets 
and string amplitudes and to derive generating functions
for the refinements of graph counting that occur in QFT. 
We have made some steps in the direction of connected graphs, 
and a more systematic understanding for the case of multiple 
external legs will be useful. In QFT, it is a familiar fact that the
generating functional of one-particle irreducible graphs is related to the generateing
functional of connected graphs by a Legendre transform.
Clarifying the implications for 
double cosets and strings will be desirable. Extending to other 
quantum field theories is an obvious direction. This would give 
new counting results for the relevant Feynman graphs, and 
could also help address some conceptual issues on the meaning 
of the string amplitude interpretation, with a view to exploring 
the possibility that the stringy combinatorics we have uncovered
is merely the tip of an iceberg, the bulk of which is a full-fledged 
QFT-string duality, which does not involve large $N$.

With regard to precise information on the asymptotic growth 
of amplitudes in perturbation theory, the counting sequences
of Feynman graphs and their asymptotics are of interest. 
For the case of vacuum graphs in $\phi^4$ theory,  
there is an asymptotic result \cite{bollobas} 
\bea 
 e^{ 15/4 } {  ( 4v)! \over  ( 4!)^v (2v)! 2^{2v} v!  }  
\eea 
For general $\phi^r$, with $v$ vertices, we need $rv $ 
to be even for non-vanishing vacuuum diagrams, i.e $rv = 2m $ 
for some $m$, and 
 the result  \cite{bollobas}  is 
\bea 
 e^{ - (r^2-1)/4 } { (2m)! \over {2^m m! r!^v v!}  } 
\eea 
We have given a string interpretation of these sequences.  
Strings at large quantum numbers often become classical. 
Can the above asymptotic result, e.g in $\phi^4$, 
 be explained by an appropriate semi-classical string ? 
For QED/Yukawa Feynman graphs or for QED, 
 with the Furry constraint, we hope that the counting sequences 
and analytic formulae we have described, along with 
techniques such as those of  \cite{bollobas}, will  allow the determination of the asymptotics.

\section*{Acknowledgements}

We thank   C. S. Chu, A. Hanany, V. Jejjala, 
T.R Govindarajan, N. Mekaryaa, K. Papadodimas,J.Pasukonis, R.Russo,
W. Spence, B. Stefanski, R. Szabo, G. Travaglini, 
P. Van Hove for useful discussions. SR 
is  supported by an STFC grant ST/G000565/1. 
RdMK is supported by the South African Research Chairs
Initiative of the Department of Science and Technology and National Research Foundation.
This project was initiated at the Mauritius 
workshop on quantum fields and cosmological inflation, June 2011, 
sponsored by NITHEP (National Institute for theoretical Physics, 
South Africa). We thank NITHEP, the Physics Department at the 
University of Mauritius, and the enthusiastic audience of students 
and faculty which contributed to a stimulating environment. We thank the 
organizers of the Corfu summer institute 2011 for an opportunity 
to present some of the results of this paper.

\begin{appendix}

\section{ Semi-direct product structure of Feynman graph symmetries.}\label{nutshell}

In this section we will explain  that the group of automorphisms of a Feynman graph can be realized
as the semi-direct product of two subgroups defined shortly. Towards this end, it is useful to recall the
definition of the semi-direct product. Given two groups $G_1$ and $G_2$, and a group homomorphism 
$\psi:G_2\to {\rm Aut}(G_1)$, the semi-direct product of $G_1$ and $G_2$ with respect to $\psi$ is denoted
$G_1\rtimes_\psi G_2$. As a set $G_1\rtimes_\psi G_2$ is the Cartesian product $G_1\times G_2$. Multiplication
is defined using $\psi$ as
\bea
  (g_1,h_1)*(g_2,h_2)=(g_1\psi_{h_1}(g_2),h_1 h_2)
\eea
for all $g_1,g_2\in G_1$ and $h_1,h_2\in G_2$. The identity element $e$ is $(e_{G_1},e_{G_2})$ and
\bea
  (g,h)^{-1}=(\psi_{h^{-1}}(g^{-1}),h^{-1})
\eea
The set of group elements $(g_1,e_{G_2})$ for a normal subgroup of $G_1\rtimes_\phi G_2$ isomorphic to $G_2$,
while the set of elements $(e_{G_1},h)$ form a subgroup isomorphic to $G_1$.

To make our discussion concrete, again consider $g\phi^4$ theory.
Recall from Section \ref{FeynPair} that  
any  graph can be   specified, after introducing a new type of vertex (say white
when the original vertices are colored black) between 
existing edges and labelling the resulting half-edges, 
by data $\Sigma_0 $ associated with the vertices and $\Sigma_1$ with the edges. 
For a graph with $v$ vertices, 
$\Sigma_0$ is a collection of $v$ 4-tuples of numbers
\bea 
\Sigma_0 = \prod_{r =1}^v \Sigma_0^{(r)} 
\eea
$\Sigma_1$ is a product of two cycles. 
The symmetric group $S_{4v}$ acts by permuting the half edges. 

We can define the Automorphism group of the graph using the data $ \Sigma_0$ and $ \Sigma_1$. It 
is the group of permutations $\gamma \in S_{ 4 v } $ which 
have the property 
\bea\label{autdef} 
&& \gamma ( \Sigma_0 ) = \Sigma_0 \cr 
&& \gamma ( \Sigma_1 ) \gamma^{-1} = \Sigma_1 
\eea
By $ \gamma ( \Sigma_0 ) $ we mean the operation which acts on, on each factor 
of $ \Sigma_0$, as follows 
\bea 
 < i ~ j ~ k ~ l > \rightarrow < \gamma ( i) ~ \gamma ( j  )  ~ \gamma ( k  ) 
~ \gamma (l )  >
\eea 
The action on $ \Sigma_1$ can also be written as above, or equivalently
in terms of conjugation in $S_{4v}$. 
In testing the first equality, we treat each angled-bracket as 
completely symmetric.

Two types of actions can be automorphisms : 
vertices (i.e. black dots) can be swapped, and propagators (i.e. white dots) can be swapped.
Construct the group $G_E$ which acts only on the propagators and the group $G_V$ that acts on the vertices. 
The elements of $G_V$ act as a non-trivial permutation on the $r$ index running over the vertices. 
The elements of $G_E$ act trivially on the $r$ index.

For any given Feynman graph we can argue that
${\rm Aut}(D)=G_E\rtimes_\psi G_V$. Towards this end
we will prove that $G_V $ acts as an automorphism of $G_E$. 
For each element $\nu \in G_V$ there is a permutation $\gamma \in S_v$
which permutes the vertices 
$$
  \nu (\Sigma_0^{(r)})=\Sigma_0^{(\gamma (r))} $$
The elements of $G_E$ leave the vertices fixed
$$
  \epsilon (\Sigma_0^{(r)})=\Sigma_0^{(r)}\qquad  \epsilon \in G_E
$$
We have 
$$
  \nu^{-1}\epsilon \nu (\Sigma_0^{(r)})=\nu^{-1}\epsilon ((\Sigma_0^{(\gamma(r))})=\nu^{-1}((\Sigma_0^{(\gamma(r))})
 =\Sigma_0^{(r)}
$$
Thus, $\nu^{-1}\epsilon\nu\in G_E$. Defining
 $\psi_\nu(\epsilon )=\nu^{-1}\epsilon\nu$, it follows that 
$\psi_{\nu_1}\psi_{\nu_2}(\epsilon)=\psi_{\nu_1 \nu_2}(\epsilon)$ which shows that we indeed have a homomorphism
$\psi :G_V\to {\rm Aut}(G_E)$.

To understand why ${\rm Aut}(D)=G_E\rtimes_\psi G_V$, consider the product law for the semi-direct product
$$
  (\epsilon_1,\nu_1)*(\epsilon_2,\nu_2)=(\epsilon_1\psi_{\nu_1}(\epsilon_2),\nu_1\nu_2)=(\epsilon_1\nu_1\epsilon_2\nu_1^{-1},\nu_1\nu_2)
$$
Acting on the graph $D$ means acting on $\Sigma_0$ and $\Sigma_1$ as defined above.
The left hand side applies $\epsilon_1\nu_1\epsilon_2\nu_2$ to $D$. The right hand side applies 
$\epsilon_1\psi_{\nu_1}(\epsilon_2)\nu_1\nu_2=\epsilon_1\nu_1\epsilon_2\nu_1^{-1}\nu_1\nu_2=\epsilon_1\nu_1\epsilon_2\nu_2$
completing the demonstration.

\begin{figure}[ht]%
\begin{center}
\includegraphics[width=0.25\columnwidth]{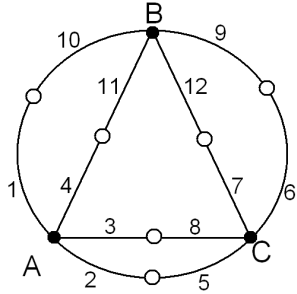}%
\caption{For this Feynman graph $v=3$.}%
\label{fig:fdiagramexample}%
\end{center}
\end{figure}

\begin{figure}[ht]%
\begin{center}
\includegraphics[width=0.5\columnwidth]{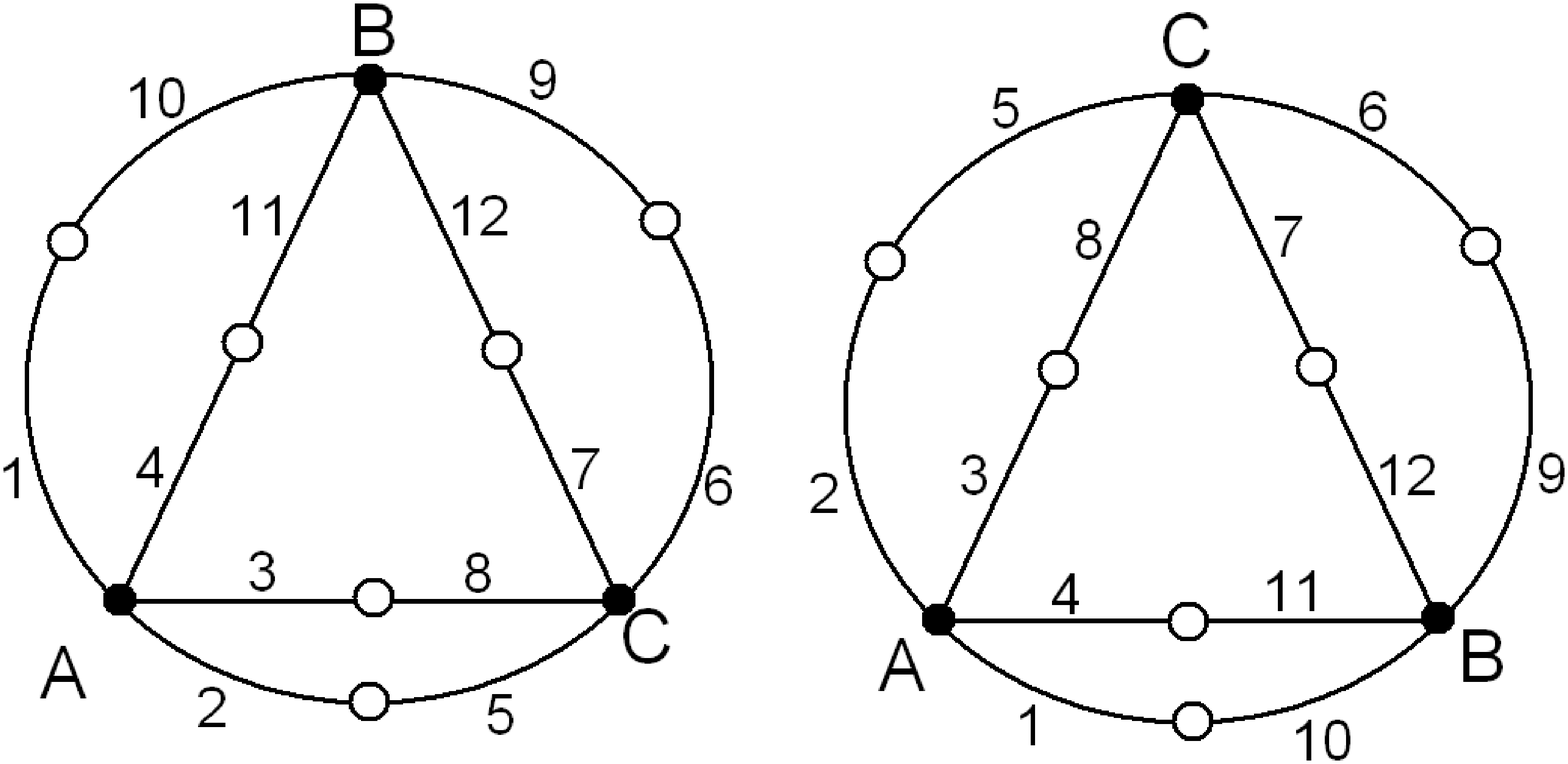}%
\caption{The two Feynman graphs shown are related by swapping vertices $B$ and $C$.}%
\label{fig:fdiagram2}%
\end{center}
\end{figure}

An example is in order.  For the graph in Figure 
\ref{fig:fdiagramexample}, we have
\bea 
\Sigma_0 & = &  < 1~ 2~ 3~ 4 > < 5~ 6~ 7~ 8>< 9 ~ 10 ~  11 ~  12 > \cr 
\Sigma_1 & = & ( 1 ~ 10 ) ( 4 ~ 11 ) ( 3 ~ 8 ) ( 2 ~ 5 ) ( 6 ~ 9 ) ( 7 ~ 12 ) 
\eea 
From the figure we read off the generators
(1 4)(10 11), (2 3)(5 8), (9 12)(6 7) for $G_E$. 
Since these generators commute, $G_E$ is a group of order
8. The group $G_V$ will permute the vertices $A$, $B$ and $C$ of the Feynman graph. There are 6 possible
permutations of the vertices. To obtain the generators of $G_V$ consider (for example) the permutation which swaps
$B$ and $C$, shown in figure \ref{fig:fdiagram2}. The relevant permutation is 
$\sigma_{BC}=$(1 2)(3 4)(5 10)(6 9)(7 12)(8 11). This is indeed an endomorphism of $G_E$ since
$$
  \sigma_{BC}(1\,\, 4)(10\,\, 11)\sigma_{BC}^{-1}=(2\,\, 3)(5\,\, 8)\qquad
  \sigma_{BC}(2\,\, 3)(5\,\, 8)\sigma_{BC}^{-1} = (1\,\, 4)(10\,\, 11)
$$
$$
  \sigma_{BC}(9\,\, 12)(6\,\, 7)\sigma_{BC}^{-1} = (6\,\, 7)(9\,\, 12)
$$
The same is true of the other elements of $G_V$. $G_V$ has order 6
and $G_E$ has order 8 so that ${\rm Aut}(D)$ is order 48. This Feynman graph thus comes with a coefficient $(4!)^3/48 = 288$.

Note that there is a notion of Graph automorphism used in graph theory
\cite{Wiki-graphauto}. This only includes $G_V$, since the standard 
labelling of graphs is to label the vertices and list the edges as pairs of 
numbers associated with the vertices they are incident on. 
The distinction is discussed in 
\cite{overflowSymms}.

\section{ Functions on the double-coset }\label{functions on double coset}  

In this Appendix we will explain how to build a complete set of 
functions on  $ S_n\setminus (S_n \times S_n )/ (H_1 \times H_2 )  $. 
This Appendix makes use of techniques similar to what was used in \cite{BHR}.

A basis of functions on $S_n$ is given by the matrix elements of 
irreducible representations. Start with an element in the group 
 algebra of $(S_n \times S_n )$ labelled by representations and states 
in the representations. 
\bea 
\tilde \cO^{R_1 , R_2}_{i_1 j_1 ; i_2 j_2 } 
= \sum_{ \sigma_1 , \sigma_2 \in S_n } 
D^{R_1}_{i_1 j_1 }  ( \sigma_1 ) D^{R_2}_{i_2 j_2 } (\sigma_2 ) 
 \sigma_1 \otimes \sigma_2 
\eea

We can make it invariant under left action of $S_n$ 
and under right action of $H_1 \times H_2$ by taking 
\bea 
 \cO^{R_1 , R_2}_{i_1 j_1 ; i_2 j_2 }  = 
{1 \over n! |H_1 | |H_2 |  }
\sum_{ \alpha  \in S_n } \sum_{ \beta_1 \in H_1 } \sum_{ \beta_2 \in H_2 } 
( \alpha \otimes \alpha )  \tilde \cO^{R_1 , R_2}_{i_1 , j_1 ; i_2 j_2 }
 ~~ \beta_1 \otimes \beta_2 
\eea
 
Some basic manipulations lead to 
\bea 
&&  \cO^{R_1 , R_2}_{i_1 j_1 ; i_2 j_2 } 
= {1 \over n! |H_1 || H_2 |  }  
 \sum_{ \sigma_1 , \sigma_2 \in S_n } 
\sum_{ \alpha  \in S_n } \sum_{ \beta_1 \in H_1 } \sum_{ \beta_2 \in H_2 }
  D^{R_1}_{i_1 j_1 }  (\alpha  \sigma_1 \beta_1 ) D^{R_2}_{i_2 j_2 } (\alpha \sigma_2 \beta_2 ) ~~ 
 \sigma_1 \otimes \sigma_2 \cr 
&& = {1 \over n! |H_1 || H_2 |  }  \sum_{ \sigma_1 , \sigma_2 \in S_n }  \sum_{ \alpha  \in S_n } \sum_{ \beta_1 \in H_1 }
 \sum_{ \beta_2 \in H_2 } \sigma_1 \otimes \sigma_2
 D^{R_1}_{i_1 k_1  } ( \alpha )  D^{R_1}_{ k_1 l_1   } ( \sigma_1  )
      D^{R_1}_{ l_1  j_1   } ( \beta_1 ) 
 D^{R_2}_{i_2 k_2  } ( \alpha )  D^{R_2}_{ k_2 l_2   } ( \sigma_2  )
      D^{R_2}_{ l_2  j_2   } ( \beta_2 ) \cr 
&& = \sum_{ \sigma_1 , \sigma_2 } \sum_{ \mu_1 , \mu_2 } 
 \sigma_1 \otimes \sigma_2 D^{R_1}_{ k_1 l_1   } ( \sigma_1  )  D^{R_2}_{ k_2 l_2   } ( \sigma_2  )\cr
&&\qquad \delta_{ R_1 , R_2 } \delta_{ i_1 , i_2 } \delta_{ k_1 , k_2 } 
B^{R_1, \mu_1 }_{ l_1  } ( { \bf 1}_{H_1} )  B^{R_1, \mu_1 }_{ j_1   } 
( { \bf 1}_{H_1} ) 
B^{R_2, \mu_2 }_{ l_2  } ( { \bf 1}_{H_2} )  B^{R_2, \mu_2 }_{ j_2  } 
( { \bf 1}_{H_2} )
\eea 
The label $\mu_1  $ runs over the multiplicity with which the 
trivial irrep. of $H_1$ appears in the decomposition of the $S_n$ 
irrep. $R$ with respect to the subgroup. Thus 

\bea  
&&  \cO^{R_1 , R_2}_{i_1 j_1 ; i_2 j_2 }
=  \delta_{ R_1 R_2 }  \sum_{ \sigma_1 , \sigma_2 } \delta_{i_1 i_2 } 
 \sigma_1 \otimes \sigma_2 D^{R_1}_{ l_1  l_2 } ( \sigma_1^{-1} \sigma_2   )
\delta_{ R_1 , R_2 }
B^{R_1, \mu_1 }_{ l_1  } ( { \bf 1}_{H_1} )  B^{R_1, \mu_1 }_{ j_1   } 
( { \bf 1}_{H_1} ) 
B^{R_2, \mu_2 }_{ l_2  } ( { \bf 1}_{H_2} )  B^{R_2, \mu_2 }_{ j_2  } 
( { \bf 1}_{H_2} )
 \cr
&& =   \delta_{ R_1 R_2 }  \sum_{ \sigma , \sigma_2   } \delta_{i_1 i_2 } 
  ( \sigma_2 \sigma^{-1}   \otimes  \sigma_2 )  \delta_{ R_1 , R_2 }
 D^{R_1}_{ l_1  l_2 } ( \sigma  )
 B^{R_1, \mu_1 }_{ l_1  } ( { \bf 1}_{H_1} )  B^{R_1, \mu_1 }_{ j_1   } 
( { \bf 1}_{H_1} ) 
B^{R_2, \mu_2 }_{ l_2  } ( { \bf 1}_{H_2} )  B^{R_2, \mu_2 }_{ j_2  } 
( { \bf 1}_{H_2} )\cr 
&& ~~ 
\eea
The sum over $ \sigma_2 $ is trivial, so it suffices to consider 
\bea 
\cO^R_{ j_1 , j_2 } 
= \sum_{ \sigma \in S_n }\sigma ~~  D^R_{l_1 l_2 } ( \sigma )  
B^{R, \mu_1 }_{ l_1  } ( { \bf 1}_{H_1} )  B^{R, \mu_1 }_{ j_1   } 
( { \bf 1}_{H_1} ) 
B^{R, \mu_2 }_{ l_2  } ( { \bf 1}_{H_2} )  B^{R, \mu_2 }_{ j_2  } 
( { \bf 1}_{H_2} )
\eea 
From this $\cO^{R_1 , R_2}_{i_1 j_1 ; i_2 j_2 }$ is reconstructed
by using $\delta_{ R_1 , R_2} \delta_{i_1 i_2 } $. 

Now define an element in the group algebra of 
$S_n$ labeled by   multiplicities $ ( \nu_1 , \nu_2 ) $ 
of the identity irrep of $(H_1, H_2 )$ appearing in a decomposition 
of  irrep $R$ of $S_n$. 
Using orthogonality of the branching coefficients we have 
\bea 
\cO^{ R  }_{  \nu_1 , \nu_2 } 
& \equiv & \sum_{ j_1 , j_2 } 
B^{R, \nu_1 }_{ j_1  } ( { \bf 1}_{H_1} )
B^{R, \nu_2 }_{ j_2  } ( { \bf 1}_{H_1} )
 \cO^{R }_{ j_1 ,  j_2 } \cr 
& = &  \sum_{ \sigma  } \sigma D^R_{l_1 l_2 } ( \sigma ) 
B^{R, \nu_1 }_{ l_1  } ( { \bf 1}_{H_1} ) B^{R, \nu_2 }_{ l_2  } ( { \bf 1}_{H_2} )
\cr 
&& ~~ 
\eea

So we see that a complete set of functions on 
the coset 
\bea 
  S_n \setminus ( S_n \times S_n ) / ( H_1 \times H_2 ) 
\eea 
can be labelled by the representations  $R$ of $S_n$
which contain the trivial of $H_1 $ and of $ H_2$. 

We  have a  complete set of elements in the group 
algebra of 
$  ( S_n \times S_n )$,  which are  invariant under the left action of $S_n$
and the right action of $ H_1 \times H_2$. 
 
This can be viewed as a derivation of the 
formula for Feynman graph counting in terms
of multiplicities given in formulas (\ref{multipls-count1}) and (\ref{multipls-count2}).

\subsection{ QED counting in terms of representation theory  }
To count the number of QED vacuum graphs with $2v$ vertices we need to evaluate 
\bea 
F_{QED} ( 2v ) && = { 1 \over 2^v v!  } \sum_{ \sigma \in S_v [ S_2 ] }  \sum_{ \gamma \in S_{2v} } 
 \delta ( \sigma \gamma \sigma \gamma^{-1} \sigma ) \cr 
&& ={ 1 \over (2v)! } 
 { 1 \over 2^v v!  } \sum_{ R \vdash 2v }  \sum_{ \sigma \in S_v [ S_2 ] }  \sum_{ \gamma \in S_{2v} } 
d_R \chi_R  ( \sigma \gamma \sigma^{-1} \gamma^{-1} ) \cr 
&& = { 1 \over (2v)! } { 1 \over 2^v v!  } \sum_{ R \vdash 2v }  \sum_{ \sigma \in S_v [ S_2 ] }  \sum_{ \gamma \in S_{2v} } 
d_R D^{R}_{ij} ( \sigma ) D^R_{jk} ( \gamma ) D^R_{kl}  ( \sigma^{-1} ) D^R_{li} ( \gamma^{-1} ) \cr 
&& = { 1 \over 2^v v!  } \sum_{ R \vdash 2v  }  \sum_{ \sigma \in S_v [ S_2 ] }\chi_R ( \sigma ) \chi_R ( \sigma )  
\eea 
We have used the orthogonality of matrix elements 
\bea 
\sum_{ \gamma \in S_{2v  } } D^R_{jk}  ( \gamma ) D^R_{li}  ( \gamma^{-1} ) = { ( 2v) ! \over d_R }  \delta_{ k l } \delta_{ ji} 
\eea
The product of characters can be be expanded using the Clebsch-Gordan (inner-product) multiplicities. 
$C(R,R, \Lambda ) $ is the number of times the representation $ \Lambda $ of $S_{2v}$ appears in 
the tensor product $ R \times R$, when this is decomposed in terms of the diagonal $S_{2v}$. Thus 
\bea 
F_{QED} ( 2v ) && =  { 1 \over 2^v v!  } \sum_{ R \vdash 2v  }  \sum_{ \sigma \in S_v [ S_2 ] }
C ( R , R , \Lambda ) \chi_{ \Lambda } ( \sigma ) \cr 
 &&   =  \sum_{ R \vdash 2v  }  C ( R , R , \Lambda )  \cM^{ \Lambda}_{ { \bf 1 }_{S_v[S_2]} } 
\eea
The multiplicity $\cM^{ \Lambda}_{ { \bf 1 }_{S_v[S_2]} }$ is the number of times the identity representation 
of $ S_v[S_2] $ appears when the representation $ \Lambda $ of $ S_{2v}$ is decomposed into the 
irreducible representations of the subgroup $S_v[S_2]$. 

Similar manipulations in the case of Furry QED leads to expressions involving, not 
Clebsch-Gordan multiplicities, but Clebsch-Gordan coefficients, along with the matrix elements 
for $ \sum_{ \tau even } \tau \otimes \tau $ in  $ R \otimes R$.

\section{ Feynman graphs with GAP }\label{FeynGAP} 

We have explained different points of view on 
the enumeration of Feynman graphs, perhaps the most intuitive 
and useful is that Feynman graph is an orbit of 
the vertex symmetry group acting on the Wick contractions. 
For the case of $\phi^4$ theory, we have the wreath product 
 $ S_v[S_4]$ acting on the conjugacy class of
permutations in $S_{4v}$ consisting of 2-cycles, which 
we denoted  $[2^{2v}] $. Calculations with this formulation 
are easy to implement directly in the GAP software 
for group theoretic computations \cite{GAP}. 

\subsection{Vacuum graphs of $\phi^4$ } 
We illustrate with the sequence  $ 1, 3, 7, 20 , 56 ... $
of vacuum Feynman graphs in $\phi^4$ theory.
The count of $3$ for $v=2$ vertices can be obtained from  GAP 
using the commands. 

\noindent
gap$>$ C $:=$ ConjugacyClass( SymmetricGroup(8), ( 1,2) (3,4) (5,6) ( 7,8)   ) ;;
 \\
gap$>$ G $:=$  WreathProduct ( SymmetricGroup(4), SymmetricGroup(2) ) ;; \\
gap$>$  Length(OrbitsDomain ( G , C ) ) ; \\
$[$ 3 

This directly implements the formulation of vacuum Feynman graphs 
as orbits of $S_v [ S_4] $ on the conjugacy class $[2^{2v}] $ 
described in section \ref{FeynPair}. The numbers $7,20$ can also be recovered 
without much trouble with this method. For the purposes of just getting
the number of vacuum graphs, this method is an overkill. Better
methods with cycle indices are described in Section \ref{sec:string}. 
We found the use of GAP to be very  useful as a tool to check 
in formulating the correct group theoretical 
formulations of various Feynman graph 
counting problems.  The command OrbitsDomain ( G , C ) actually 
gives a nested list of lists of  pairings. The list runs 
over inequivalent Feynman graphs. For each Feynman graph, there is 
a list of different Wick contractions which lead to the given 
graph. Hence these orbits encode not just the number of Feynman graphs, 
but the Feynman graphs themselves. 
 
\subsection{ $\phi^4$ with external edges } 

It is a small generalization to consider graphs with external edges.
For $\phi^4$ theory graphs with $E$ external edges we act with the wreath
product $S_v [S_4]$ on the conjugacy class $[2^{2v+{E\over 2}}]$ of
$S_{4v+2E}$. The relevant sequence for $E=2$ is $1, 2, 7, 23, 85, 340,...$.
The count of $7$ for $v=2$ and $E=2$ can be obtained from GAP as follows

\noindent
gap$>$ C $:=$ ConjugacyClass( SymmetricGroup(10), ( 1,2) (3,4) (5,6) (7,8) (9,10)  ) ;; \\
gap$>$ G $:=$  WreathProduct ( SymmetricGroup(4), SymmetricGroup(2) ) ;; \\
gap$>$  Length(OrbitsDomain ( G , C ) ) ; \\
$[$ 7

Comments made above are again applicable: there are better methods to count these
graphs which use cycle indices.

\subsection{Symmetry factor } 
The symmetry factor of a given Feynman graph can be constructed
using GAP. The following commands compute the symmetry factor 
of the diagram in Figure \ref{fig:fdiagramexample}. 

\noindent 
gap$>$  H1 $:=$  WreathProduct( SymmetricGroup(4) , SymmetricGroup(3) ) ; \\
gap$>$  Sig1 $ := $  (1,10) ( 4,11) (3,8) (2,5) (6,9) ( 7,12) ; \\
gap$>$  for g in H1 do \\ 
  $ ~~~~~~  ~~~~  $   if OnPoints( Sig1 , g ) $:= $  Sig1 then\\ 
    $ ~~~~~~~~~ $    Countsym := Countsym +1 ; fi ; od ; Countsym ; \\
$[$ 48 

Simple modifications of these commands can construct the group, 
study its subgroups etc. , but these will have no immediate interest for us.

\subsection{Action of $S_v$ on a set  }
 
An alternative way to code the graphs is to 
label the vertices $ \{ 1 , \cdots , v \} $. 
Consider lists of $ 2v$ unordered pairs. 
\bea 
\{  ( i_1, j_i ) , ( i_2 , j_2 ) ... ( i_{2v} , j_{2v } ) \} 
\eea
Put the constraint that 
\bea 
\sum_{k=1}^{2v} \delta ( i , i_k ) + \delta ( i , j_k ) = 4 
\eea
for all $i$ from $1, .. v$. This imposes the condition that all vertices are 4-valent. 

There is an action of $S_{v}$ on this set. We are interested in the orbits. 
This can be programmed in GAP  and gives an alternative  method 
to get the sequence of vacuum Feynman graphs in $\phi^4$ theory.

The method of using $S_{v}$ is more efficient  in generating the
 whole set of Feynman graphs 
 with GAP, than the $ S_v [S_4] $ method. But it does not 
automatically  have the edge symmetries, unlike the 
$  S_v [S_4] $ approach. It only gives the automorphism $G_V$ 
discussed in Appendix \ref{nutshell}.

\section{ Integer sequences  }\label{data} 

In this Appendix we collect the numerical results we have obtained by counting Feynman graphs.

\begin{itemize}

\item The number of vacuum graphs in $ \phi^4$ theory, with $v$ vertices, starting with $v=1$, is
      given by the sequence 
      \bea 
      &&1, 3, 7,  20, 56, 187, 654, 2705, 12587, 67902, 417065, 2897432, 22382255, 189930004,\nonumber\\
      &&~1750561160, ..
      \eea
      This sequence is listed in ``The On-Line Encyclopedia of Integer Sequences'' \cite{OEIS}, where it is described 
      as the sequence of 4-regular multi-graphs (loops allowed). 
      This sequence is derived using (5.21) or (6.4). 

\item The number of connected vacuum graphs in $\phi^4$ theory with $v$ vertices, starting with $v=1$, is given by 
      the sequence
      \bea
      &&1,~ 2,~4,~10,~28,~97,~359,~1635,~8296,~48432,~316520,~2305104,\nonumber\\
      &&~18428254,...
      \eea
      This sequence was generated using (\ref{frstconnected}).

\item The number of graphs in $ \phi^4$ theory with $E=2$ external legs, with $v$ vertices, starting with $v=0$, is
      given by the sequence 
      \bea 
      &&1,~2,~7,~23,~85,~340,~1517,~7489,~41276,~252410,~1706071,~12660012,\nonumber\\
      &&~102447112,...
      \eea
      This sequence is generated by the formula (\ref{Eistwo}) with $E=2$. 

\item The number of connected graphs in $\phi^4$ theory with $E=2$ external legs, with $v$ vertices, starting with $v=0$, is
      given by the sequence 
      \bea 
      &&1,~1,~3,~10,~39,~174,~853,~4632,~27607,~180148,~1281437,~9896652,\nonumber\\
      &&~82610706,...
      \eea
      This sequence is generated by the formula (\ref{scndconected}). These graphs have been tabulated in \cite{Kastening}.
      Comparing the Table II of \cite{Kastening}, we find a match between the number of diagrams at first and second order
      in perturbation theory. At third order in perturbation theory we have 10 graphs compared to the 8 graphs listed in \cite{Kastening}.
      The two graphs not listed in \cite{Kastening} are obtained from graph \# 5.3 by swapping the position
      of the closed loop and setting sun, and from \# 7.2 by swapping the order of the double bubble and the bubble. At fourth
      order in perturbation theory, there are 30 graphs listed in \cite{Kastening} compared to our count of 39. However, there
      are 9 graphs that are obtained from graphs \#11.2, \#11.4, \#12.3, \#14.2, \#14.3, \#15.1,  \#13.3, \#15.5 and \#17.2, in much
      the same way that we described for third order in perturbation theory.       

\item The number of graphs in $ \phi^4$ theory with $E=4$ external legs, with $v$ vertices, starting with $v=0$, is
      given by the sequence 
      \bea 
      &&3,~10,~44,~190,~889,~4490,~24736,~148722,~976427,~6980529,~54151689,\nonumber\\
      &&~453922676,...
      \eea
      This sequence is generated by the formula (\ref{Eistwo}) with $E=4$. 

\item The number of vacuum graphs for $\phi^3$ theory, with $2v$ vertices, starting from $v=1$
      \bea
      2,~8,~31,~140,~722,~4439,~32654,~289519,...
      \eea
      This sequence was generated using (\ref{vacuumphi3}).

\item The number of connected vacuum graphs in $\phi^3$ theory, with $2v$ vertices, starting from $v=1$
      \bea
      2,~5,~17,~71,~388,~2592,~21096,~204638,...
      \eea 
      This sequence was generated using (\ref{frstconnected}).

\item The number of graphs in $\phi^3$ theory with $E=2$ external legs, with $2v$ vertices starting from $v=1$, is
      \bea
      5,~30,~186,~1276,~9828,~86279,~866474,~9924846,...
      \eea
      This sequence was generated using (\ref{extlegsphi3}) with $E=2$.

\item The number of QED/Yukawa vacuum graphs with $2v$ vertices, which equals the total number of ribbon graphs 
      with $2v$ edges is given by 
      \bea
      2,~8,~34,~182,~1300,~12634,~153590,~2230979,~37250144,...
      \eea
      This sequence was generated using (\ref{QEDCOUNT}).

\item The number of connected QED/Yukawa vacuum graphs with $2v$ vertices, starting from $v=1$, is given by 
      \bea
      2,~5,~20,~107,~870,~9436,~122832,~1863350,~32019816,...
      \eea
      This sequence was generated using (\ref{frstconnected}).

\item Vacuum graphs in QED after implementing the constraint due to Furry's theorem. 
      \bea
       1,~4,~12,~57,~321,~2816,~31092,~423947,... 
       \eea
       This sequence was generated using (\ref{FurryQED}).

\item Connected Vacuum graphs in QED after implementing the constraint due to Furry's theorem. 
      \bea
       1,~3,~8,~39,~240,~2332,~27196,~382,~802,...  
       \eea
       This sequence was generated using (\ref{frstconnected}).
       The $1,3,8$ agree with the results of \cite{cvitan2}.

\end{itemize}

\end{appendix}

\end{document}